\documentclass[preprint]{aastex}
\input epsf
\newcommand\be{\begin{equation}}
\newcommand\ee{\end{equation}}

\newcommand\aaa{{\bf a}}
\newcommand\bb{{\bf b}}

\newcommand\kk{{\bf k}}

\def\r{{\bf r}}
\newcommand\s{{\bf s}}
\def\v{{\bf v}}

\newcommand\x{{\bf x}}

\newcommand\z{{\bf z}}

\newcommand\Alfven{Alfv\'en }

\newcommand\hh{\hspace{1mm}}

\newcommand\tablemethod {1}
\newcommand\tablemethodz{2}

\begin{document}

\title{\mbox{Gradient Particle Magnetohydrodynamics}}

\author{Jason L. Maron\altaffilmark{1} and Gregory G. Howes\altaffilmark{2}}
\affil{Department of Physics and Astronomy, University of California,
    Los Angeles, CA 90095-1547}
\altaffiltext{1}{maron@tapir.caltech.edu}
\altaffiltext{2}{ghowes@physics.ucla.edu}

\begin{abstract} 
We introduce Gradient Particle Magnetohydrodynamics (GPM), a new
Lagrangian method for magnetohydrodynamics based on gradients
corrected for the locally disordered particle distribution.  The
development of a numerical code for MHD simulation using the GPM
algorithm is outlined.  Validation tests simulating linear and
nonlinear sound waves, linear MHD waves, advection of magnetic
fields in a magnetized vortex, hydrodynamical shocks, and
three-dimensional collapse are presented, demonstrating the viability
of an MHD code using GPM. The characteristics of a GPM code are
discussed and possible avenues for further development and refinement
are mentioned.  We conclude with a view of how GPM may complement
other methods currently in development for the next generation of
computational astrophysics.
\end{abstract}

\section{Introduction} Computer modeling of astrophysical fluids often
requires the accurate representation of densities and other fluid
quantities which vary over several orders of magnitude due to the
inherent compressibility of the interstellar medium. This challenge
has often been met by the use of Lagrangian particle methods to
simulate astrophysical fluid flow.  The ``particles'' in a simulation
represent fluid elements. When the fluid is compressed to high
densities, the particles---points where we know information about the
fluid---flow with the fluid, resulting in increased resolution in the
dense regions.  The relative computational ease with which
resolution is enhanced in a regions of increased density has made
Lagrangian methods very attractive to astrophysicists.  Grid-based
methods entail great computational complexity in order to attain such
a selective resolution enhancement.  The drawback of Lagrangian
methods is that it is more difficult to achieve the desirable
conservation properties characteristic of grid-based methods.

Smoothed Particle Hydrodynamics (SPH) is a Lagrangian technique that
has seen widespread use since its introduction by \citet{luc77} and
\citet{mon77} two decades ago.  Although the technique does not
provide a solution to high accuracy, SPH has proven extremely valuable
through its ability to yield solutions to many problems that other
computational methods could not possibly tackle.

Today, computational astrophysicists are seeking to extend the limits
of applicability of their techniques.  Those using grid-based methods
have turned to Adaptive Mesh Refinement (AMR) to push the limits of
these high-accuracy methods to allow larger variation in density. For
those employing SPH codes, the inclusion of magnetic fields has become
a priority to apply the method to a wider range of phenomena in which
the dynamical effect of magnetic fields cannot be neglected.

Gradient Particle Magnetohydrodynamics (GPM) is a new algorithm
introduced here for Lagrangian simulation of magnetohydrodynamics
(MHD).  It is, essentially, an algorithm for correctly calculating the
gradient of fluid quantities in the presence of particle disorder.
SPH, a Monte Carlo technique, fails to stably include magnetic fields
because of the small-scale noise inherent in Monte Carlo methods.  GPM
determines the gradients of fluid quantities exactly, rather than
statistically, and therefore is not susceptible to the same magnetic
tension instability arising in SPH.

We begin, in Section~\ref{sec:problems}, with a discussion of the
problems with existing particle methods for hydrodynamics and describe
the GPM algorithm which solves these problems.  In
Section~\ref{sec:MHD}, we describe the application of the GPM
algorithm to the equations of MHD and discuss viscosity, magnetic
divergence, and advanced features to be implemented in the code.
Analytical estimates of the error of this method are presented in
Section~\ref{sec:convergence}.  Section~\ref{sec:valid} presents the
validation tests performed with the new GPM technique.  Issues arising
from the validation tests are discussed in Section~\ref{sec:disc} and
concluding remarks are made in Section~\ref{sec:conc}.

\section{Properties of Particle Hydrodynamics}
\label{sec:problems}

Lagrangian numerical methods for hydrodynamics face the difficult task
of computing fluid forces accurately when information about the fluid
is known only at a discrete set of points whose positions and number
may vary.  Existing techniques work well when the particles are
relatively ordered; problems occur, however, when the particles become
disordered (as they often do). We present below a simple example in
which this problem is apparent; we then present the GPM algorithm as
a solution to this problem.

\subsection{Difficulty with Particle Disorder}
A fluid is modeled as a collection of particles whose positions need
not fall on a regular lattice, and where dynamical forces are computed
by sampling over neighboring particles \citep{mon85}.  Let $q$ be an
arbitrary fluid quantity. The mean and gradient (at $\r=0$) are
calculated by convolving neighboring particles $i$ with a symmetric
smoothing kernel $W(\r,h)$ and a gradient kernel $x_j W(\r,h)$:
\begin{equation} <q>
= \frac{\sum_i q W(\r)}{\sum_i W(\r)}, \hspace{15mm} <\partial_j q> =
\frac{\sum_i q x_j W(\r)}{\sum_i x_j^2 W(\r)}. 
\end{equation} 
The characteristic smoothing radius $h$ of $W(\r,h)$ is arranged to
include enough neighbor particles to sample the local environment
adequately, otherwise the precise form of the kernel is a matter of
engineering.

\begin{figure}[h!]
\hbox to \hsize{ \hfill \epsfxsize6cm
\epsffile{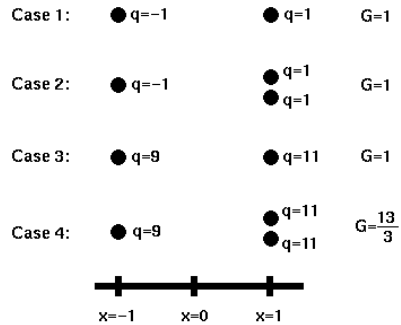} \hfill }
\caption{ Gradient noise from an irregular particle
distribution.}
\label{fig:disorder}
\end{figure}

A spatially irregular particle distribution confounds the calculation
of gradients, as is illustrated by the 1-dimensional test cases in
Figure~\ref{fig:disorder}.
For this example, let $W(x) = 1$ if $|x| \le 1$ and $0$
otherwise.  In cases 1 and 3 the particles are regularly distributed:
one each at $x=+1$ and $-1$. The particles in cases 2 and 4 are
irregularly distributed with two at $x=1$ and one at $x=-1$. In every
case, the true gradient $\partial_x q$ is equal to $1$, but in cases 1
and 2 the average value of $q$ is $0$ and in cases 3 and 4 it is
$10$. The gradient operator $G$ yields the correct value in every case
except number $4$.  Here, the background value of $q$ introduces a
gradient noise which obliterates the true gradient. Additional
measures must be taken to extract gradients in the presence of a
background. It may also be noted that the irregular particle
distributions in cases 2 and 4 disrupt the evaluation of $<\!q\!>$.

SPH employs a similar procedure to evaluate pressure gradients.  The
irregular particle distribution gives rise to artificial fluctuations
in the local density and pressure with a fractional magnitude of 1,
even in a globally uniform fluid. The resulting pressure gradients
give rise to Mach 1 fluctuating particle velocities.  The physical
velocity field may be obtained by spatially averaging.  However, for a
subsonic situation, a large number of particles must be averaged to
yield a smooth flow. Therefore, the resolution per particle is quite
low.

SPH correctly captures the physics of fluid turbulence in spite of the
loss of resolution from pressure gradient noise, however the inclusion
of magnetic fields is inviable. Consider a uniform magnetic field in a
stationary fluid. The gradient noise in the induction equation will
quickly produce small scale fluctuating fields with the same magnitude
as the uniform field. These will give rise to extreme forces and hence
even more magnetic fluctuations, resulting in instability.

GPM suppresses noise in the pressure and magnetic gradients, promoting
reduced particle velocity fluctuations and enabling the inclusion of
magnetic fields. An additional measure may be taken to quell noise in
subsonic flows. Here, the physical density fluctuations have a
magnitude of $\Delta\rho/\rho \sim M^2$, where $M$ is the Mach number.
If the density is evaluated from local smoothing, an irregular
particle distribution adds fractional density fluctuations of order
unity, triggering Mach number unity pressure and velocity
fluctuations. This can be corrected by regarding density as a particle
property which is evolved according to the continuity equation.

\subsection{Gradient Particle Magnetohydrodynamics}

The GPM gradient evaluation corrects for the irregular particle distribution.
We first consider the 1-D case. Assume that a quantity $q$
has a spatial profile of $q = A_0 + A_1 x$ and evaluate the quantities
\begin{equation}
Q_0 = \sum_i q m W = A_0 \sum_i m W
+ A_1 \sum_i m x W
\label{disa}
\end{equation}
\begin{equation}
Q_1 = \sum_i q m x W = A_0 \sum_i m x W + A_1 \sum_i m x^2 W
\label{disb}
\end{equation}
with the sum occurring over neighboring particles inside the
smoothing sphere. The smoothing kernel is $W(r) = exp(-4 r^2/h^2)$,
where $h$ is the smoothing length and $m$ is the mass of the neighbor
particle.
If the particles are symmetrically distributed, the $A_1$ term in
(\ref{disa}) and the $A_0$ term in (\ref{disb}) are
zero, and
\begin{equation}
A_0 = \frac{\sum_i q m W} {\sum_i m W} \hspace{15mm}
A_1 = \frac{\sum_i q m x W} {\sum_i m x^2 W}
\end{equation}
These formulae are equivalent to the SPH evaluation.  If the particle
distribution is irregular, the $2\times2$ matrix may be solved to
obtain $A_0$ and $A_1$.  For three dimensions, assume $q = A_0 + A_1
x_1 + A_2 x_2 + A_3 x_3$ and solve the $4\times4$ matrix. GPM can be
further extended to second order by solving the $10\times10$ matrix
resulting from $q = A_0 + A_1 x_1 + A_2 x_2 + A_3 x_3 + A_{11} x_1^2 +
A_{22} x_2^2 + A_{33} x_3^2 + A_{12} x_1x_2 + A_{13} x_1x_3 + A_{23}
x_2x_3$. Second order allows the viscous and resistive terms to be
evaluated directly.

\subsubsection{Convergence}\label{sec:convergence}

GPM quantities can be evaluated to a hierarchy of orders.  Consider
the calculation of a 1D gradient of $q$ where $q = A_0 + A_1 x + A_{11}
x^2 + \cdots$. We have
\begin{equation} Q_1 = \sum_i q x
W = A_0 \sum_i q x W + A_1 \sum_i q x^2 W + A_{11} \sum_i q x^3 W
+ \cdot\!\cdot\!\cdot.
\end{equation}
The magnitudes of the RHS terms for an
irregular particle distribution are $A_0 h n^{1/2} + A_1 h^2 n +
A_{11} h^3 n^{1/2} + \cdots$, where $h$ is the smoothing
length and $n$ is the number of particles inside the smoothing
sphere. Terms of odd order in $h$ have zero mean, and for them we
specify the fluctuating magnitude. Terms of even order in $h$ have a
positive mean with the specified magnitude. In the evaluation of
$A_1$, the $A_0$ and $A_{11}$ terms constitute the error. The fractional
error due to $A_0$, $n^{1/2} A_0/ (h A_1)$, is unbounded,
necessitating the simultaneous evaluation of $A_0$ and $A_1$. The
fractional error due to $A_{11}$ is $n^{-1/2} h A_{11} / A_1$. An
imposed viscosity can ensure that $A_{11} h$ is less than the RMS
value of $A_1$, and so the $A_{11}$ term can be neglected when
evaluating $A_{1}$. Similarly, the evaluation of $A_{11}$ requires
$A_0$ and $A_{1}$ and not $A_{111}$ because a sufficiently smooth
profile will have $A_{111} h < A_{11}(RMS)$.


\section{Application of GPM to MHD}
\label{sec:MHD}
The GPM algorithm is essentially a recipe for correctly computing
gradients in a Lagrangian fluid code.  In this section, we describe
the application of the GPM algorithm to create a working code for MHD
simulation.  We first discuss the basic application to the most simple
MHD system neglecting viscosity and resistivity, describing the kernel
used and the spatial and temporal order of the method.  Next we
describe the incorporation of viscosity into the code and suggest a
method for the elimination of magnetic divergence.  Finally, advanced
features to improve the method are described.

\subsection{Basic Application}

The governing equations of MHD are the momentum equation, the
induction equation, the continuity equation, and the energy equation:

\begin{equation}
d_t v = - \frac{1}{\rho} \nabla P
+ \frac{1}{4 \pi \rho}(\nabla \times b) \times b
+ \nu \nabla^2 v \end{equation}
\begin{equation}
d_t b = b \cdot \nabla v - b \nabla \cdot v + \eta \nabla^2 b \end{equation}
\begin{equation}
d_t \rho = - \rho \nabla \cdot v
\end{equation}
\begin{equation}
d_t e = -\frac{P}{\rho} \nabla \cdot v.
\end{equation}
The system of MHD equations is closed using the adiabatic equation of state
\begin{equation}
P = (\gamma-1) \rho e.
\end{equation}
\begin{center}
\begin{tabular}{llll}
\hline
Quantity          &Symbol     & Quantity & Symbol \\
\hline
Velocity          &$v$        & Magnetic field          &$b$ \\
Density           &$\rho$     & Pressure                &$P$ \\
Energy density    &$e$        & Ratio of specific heats &$\gamma$\\
Viscosity         &$\nu$      & Resistivity             &$\eta$\\
\hline
\end{tabular} \end{center} \vspace{0.1 cm}
In all our tests, we have chosen the 3-D adiabatic index for a
monatomic gas, $\gamma=5/3$.

We begin with an ideal MHD system, neglecting viscosity
and resistivity.  To apply the GPM algorithm to the governing
equations, it is necessary to choose the form of the kernel
$W(\mathbf{r},h)$ and the smoothing radius $h$ over which the kernel
applies.  

For a test particle at position $\r$ and a neighbor at position $\r'$,
we choose a Gaussian kernel of the form
\begin{equation}
 W( \mathbf{r}-\mathbf{r}',h)= \frac{1}{N} e^{\frac{-4(
 \mathbf{r}-\mathbf{r}')^2}{h^2}}
\end{equation}
for $\mathbf{r}-\mathbf{r}'\le h$.  We cut off the Gaussian at $r=h$,
so that for $\mathbf{r}-\mathbf{r}'>h$, $W(
\mathbf{r}-\mathbf{r}',h)=0$. The normalization factor $N$ must be
chosen such that the kernel satisfies the normalization condition
\begin{equation}
 \int_{-\infty}^{\infty} W( \mathbf{r}-\mathbf{r}',h) d\mathbf{r}' =1. 
\end{equation}
For a Gaussian truncated at $r=h$, the normalization factor is thus
given by
\begin{equation}
N= h^3 \pi^{3/2} [\mathrm{Erf}(2)]^3.
\end{equation}

All particles within a radius $h$ of the test particle will contribute
to the MHD forces on the test particle with the weight of each particle's
influence given by smoothing kernel.  An appropriate choice of
smoothing length $h$ must be made in order to include enough neighbors
within the smoothing sphere to yield a good sampling of the local
fluid characteristics, and thus calculate gradients accurately.  To
satisfy this condition, the minimum number of neighbors necessary is
approximately 12 neighbors in 2-D simulations and  32
neighbors in 3-D.  This estimate is made assuming that particles
\emph{do} lie on a uniform grid and that $h \simeq 2s$, where $s$ is
the interparticle separation.  In practice, if the particles do become
irregularly distributed, a greater number of neighbors should be
included in the smoothing sphere to insure that local fluid conditions
are sampled adequately.

The order of the GPM method is specified according to the computing
resources available and the desired accuracy of the solution.  In 3-D,
a first order GPM calculation requires the inversion of a $4 \times 4$
matrix, and a second order calculation requires a $10 \times 10$
matrix to be inverted.  We accomplish the inversion using LU
Decomposition.  It is worthwhile to note that, for each fluid quantity
whose gradient is to be calculated--- \emph{i.e.} pressure or a component of
velocity---the determination of the lower and upper triangular
matrices needs only be done once per particle.  Any gradient desired
for that particle is calculated by backsubstitution using the same
decomposed matrices.

The timestepping employed for these tests was a simple Eulerian
first-order scheme.  

In order to avoid unphysical fluctuations in the density, the mass
density $\rho$ was evolved entirely as a particle characteristic
rather than being coupled to the local number density of particles.
Hence, our ``particles'' are really not physical entities at all but
simply positions where we know information about the fluid.  This
simplifies the setting of initial conditions and also allows the
resolution to be enhanced in a particular region by simply placing
more particles in that region.  This freedom is useful because you do
not \emph{a priori} desire less resolution in a region which has a
lower density.  It does, however, allow for the mass density to move
with respect to the particle points in our simulation.  If this causes
difficulties to arise, they can be alleviated by particle removal and
replacement as will be discussed in Section~\ref{sec:advfeat}.

The basic implementation of GPM for MHD as described thus far was
susceptible to a slowly growing smoothing length related instability.
Hence, the addition of viscosity was found to be necessary to yield a
stable scheme.

\subsection{Viscosity}
Lagrangian codes typically have a very small diffusivity when compared
to grid-based methods for the same problem.  We have found that GPM,
since it removes a significant component of noise present in SPH
codes, is even more non-dissipative. The addition of artificial
velocity was found necessary to stabilize a slowly growing smoothing
length related instability and to prevent particle interpenetration in
the presence of shocks.  We investigated the different forms of
artificial viscosity and also explored the possibility of using real
viscosity and resistivity in the case of second order GPM when the
Laplacian of velocity $\v$ and magnetic field $\bb$ can be calculated
directly.

\subsubsection{Artificial Viscosity in GPM}
A common treatment of artificial viscosity in finite difference
calculations involves the addition of a bulk viscosity, which enhances
the pressure when $\nabla \cdot \mathbf{v} < 0$ \citep{roa75}. In the
momentum equation, the pressure $P$ is replaced by $P+q$, where
\begin{equation}
q = \left\{ \begin{array} {ll} -\alpha \rho h c_s \nabla \cdot
\mathbf{v} + \beta \rho h^2 (\nabla \cdot \mathbf{v})^2 & \mbox{if
$\nabla \cdot \mathbf{v}<0$} \\ 0 & \mbox{if $\nabla \cdot
\mathbf{v}>0$}.  \end{array}
\right.
\label{eq:viscpress}
\end{equation} 
Here $\alpha$ and $\beta$ are dimensionless constants, $h$ is the cell
width, $\rho$ is the local density, and $c_s$ is the local sound
speed.

\citet{mon83} suggested that for SPH, which is significantly
less diffusive than grid-based methods, artificial viscosity is always
necessary but that the above formulation smears out shock fronts
excessively because $\nabla \cdot \mathbf{v}$ is averaged over all
particles in a smoothing radius.  They found a more effective
artificial viscosity based on interparticle velocity differences
\citep{mon85,mon92}.

We have used a similar approach to \citet{mon92}, estimating $\nabla
\cdot \mathbf{v}$ by the velocity differences between particles and
enhancing the pressure of approaching particles by $q$ as in
equation~(\ref{eq:viscpress}).  In one-dimension, we estimate by
Taylor expansion the velocity divergence between a test particle $i$ and
its neighbor particle $j$ as
\begin{equation}
 \frac{(v_{xi} -v_{xj})(x_i-x_j)} {(x_i-x_j)^2} 
\simeq \frac{1}{x_i-x_j} \left\{ v_{xi} - \left[ v_{xi} +(x_j -x_i) 
\left. \frac{\partial v_{x}}{\partial x} \right|_{x_i}
+ \cdots \right] \right\} 
\simeq \frac{\partial v_x}{\partial x}.
\end{equation}
In practice, we approximate the 3-dimensional interparticle divergence as
\begin{equation}
\nabla \cdot \mathbf{v} \simeq \frac{ (\mathbf{v}_i-\mathbf{v}_j) \cdot 
(\mathbf{x}_i-\mathbf{x}_j)}{(\mathbf{x}_i-\mathbf{x}_j)^2 
+ (\epsilon h)^2/4 }
\label{eq:divv}
\end{equation}
where $\epsilon =0.1$ and $h$ is the GPM smoothing radius.  The
extra term in the denominator prevents the singularity as
$\mathbf{x}_i-\mathbf{x}_j \rightarrow 0$. The modified pressure $P+q$
is given by equation~(\ref{eq:viscpress}) with $\nabla \cdot
\mathbf{v}$ estimated by equation~(\ref{eq:divv}).  Thus, the
pressure from each neighboring particle is enhanced by the viscous
term only when it is approaching the test particle.  The modified
pressure $P+q$ is then used in the standard GPM implementation to
find the gradient of the pressure in the momentum equation
\begin{equation}
d_t \mathbf{v} = -\frac{1}{\rho} \nabla (P+q) + \frac{1}{4 \pi \rho} (\nabla
\times \mathbf{b}) \times \mathbf{b}.
\label{eq:momvisc}
\end{equation}

The artificial viscosity of the form in equation~(\ref{eq:viscpress})
behaves as a normal bulk viscosity plus a von Neumann-Richtmeyer bulk
viscosity.  For linear problems in which the stabilizing effects of an
artificial viscosity are desired, but the dissipation is to be kept to
a minimum, values of $\alpha =0.05$ and $\beta =0.1$ have proved
sufficient to stabilize smoothing length related computational
instabilities but have not altered the waveform.  For shock capturing,
larger values of $\alpha =0.5$ and $\beta =1.0$ can prevent
particle interpenetration and damp post-shock oscillations.

\subsubsection{Real Viscosity and Magnetic Diffusivity}
The ability for second order GPM to yield not only the gradient but
also the second derivative of fluid quantities opened up the potential
for employing a real viscosity and real resistivity in simulations.
Initial tests with real viscosity, however, have demonstrated a slowly
growing instability which prevented long-time simulations from being
run.

\subsection{Magnetic Divergence}

GPM does not preserve magnetic divergencelessness, however divergences can
be removed with a procedure analogous to a gravitational potential solution.
Solve $\nabla^2 \Phi = - \nabla \bb$ for $\Phi$ and reset the
magnetic field to $\bb_{\mathrm{new}} \leftarrow \bb_{\mathrm{old}} +
\nabla \Phi$.  The solution of Laplace's equation may be piggy-backed
with the $N \log_2 N$ tree gravitation algorithm.

\subsection{Advanced Features}
\label{sec:advfeat}
The most simple advanced feature that has been implemented with the
GPM is a variable smoothing length $h$. By adjusting the smoothing
length to encompass a specified number of nearest neighbors, we can
tackle a problem involving multiple density scales without excessive
smoothing in high density regions or undersampling of nearest
neighbors in the low density regions. In practice,we choose an optimal
number of neighbors, \emph{i.e.} 45 neighbors for a 3-D calculation,
and allow a range of $\pm 33\%$ from that number. We have implemented
this featuring for testing and have noted in Section~\ref{sec:valid}
when a variable smoothing length has been used for a test run.

Because the mass density $\rho$ is not fixed to the ``particle''
points, it is possible over many timesteps for the mass to slip with
respect to the particles.  Two possible problems can result: two
particles may move very close together and effectively reduce the
resolution of the simulation by including a virtually redundant point,
or particles may move away from each other leaving a region of the
fluid that is poorly sampled.  Particle removal and replacement can
solve this problem.  Particle removal has been implemented and has
proven useful in the case of collapse in a fixed central gravitational
potential.  When two particle are separated by a distance of
$r<h/100$, one particle is removed and the fluid quantities (including
position) are combined to conserve mass, center of mass, momentum,
energy, and magnetic flux.  Particle replacement in poorly sampled
regions has not yet been implemented.

The simulation of a galactic disk with magnetic fields presents severe
obstacles to any numerical technique. Turbulent structure exists at
widely varying space and time scales, rotation times vary widely with
radius, and magnetic fields and gravity are significant. For example,
molecular cloud and supernova dynamics occur at substantially smaller
scales than that of the disk. Also, the ISM consists of a mixture of
phases with widely varying temperature and density.  For a grid-based
code, the timestep is determined by the fastest and smallest scales in
the system.  If these scales occupy a small fraction of the volume,
the timestep is too small for the majority of the system. A
cylindrical coordinate system co-rotating with the inner radii can in
part compensate for the range of radial dynamics, but it cannot
simultaneously offer high resolution anywhere but at the inner radii.

We plan to incorporate the particle-based GPM fluid algorithm with a
Barns-Hut tree to efficiently handle situations with widely varying
space and time scales. In this code, particles have independent
timesteps and smoothing lengths which adjust to local
conditions. Timesteps are Lagrangian, as opposed to Eulerian, which
means they are a function of local velocity dispersion, as opposed to
the global sweeping velocity.  Gravitational forces and near-neighbor
lists for GPM are computed simultaneously with a Barnes-Hut tree, an
$N \log_2 N$ operation.  The tree is rebuilt once every 16 timesteps,
and this operation constitutes a negligible fraction of the
computational time.  The program is MPI parallel, with communication
occurring between nodes only at the start of a tree rebuild.  Memory
access is not a factor in execution speed, which is accomplished by
organizing the data by particle, linear in memory. Nearby particles in
space are stored nearby in memory to minimize cache misses.  Finally,
data is prefetched from RAM in advance of use, with the prefetch
occuring simultaneously with other floating point operations.


\section{Validation}
\label{sec:valid}

In this section, we present the results of a suite of validation tests
to determine the performance of the GPM algorithm in simulating MHD
phenomena.  We tested the propagation of linear and nonlinear sound
waves and determined a dispersion relation for varying spatial
resolution.  Sound wave test results are compared with SPH results
using the publicly available Hydra code by \citet{cou95}.  Slow,
\Alfven, and fast MHD waves were tested at the full range of angles
between the direction of propagation $\kk$ and the magnetic field
$\bb$.  A polar plot of MHD wave velocities shows excellent agreement
with theory at moderate resolution.  An advective MHD test is
performed by initializing a magnetized vortex and following the
evolution of the particles and the magnetic field; good agreement was
found with results from a spectral MHD code by \citet{mar01}. The
standard Sod shock test \citep{sod78} was performed to determine the
shock-capturing ability of GPM.  Finally, a 3-D collapse problem was
run to demonstrate the multiscale capability of GPM using variable
smoothing lengths.

All tests were run with a minimum of two dimensions, since a lot of
the problems that arise using SPH disappear when it is run in only one
dimension; in all cases, no differences were seen between
two-dimensional and three-dimensional GPM simulations when run with
the same parameters and initial conditions.  In all plots shown here,
all redundant points are plotted to demonstrate the minimal
cross-field dispersion characteristic of GPM.  All tests were run
using the adiabatic index appropriate for 3-D gas dynamics,
$\gamma=5/3$.

Clearly additional refinement of the GPM method is possible, but these
tests demonstrate the validity of the method in simulating subsonic
and supersonic flows with magnetic fields over varying spatial scales.

\subsection{Sound Waves}
To test the ability of GPM to handle hydrodynamics accurately, we
performed tests with linear and nonlinear sound waves and
determined the dispersion relation of the method when the resolution
is varied.

\begin{figure}[h!]
\hbox to \hsize{ \hfill \epsfxsize8cm \epsffile{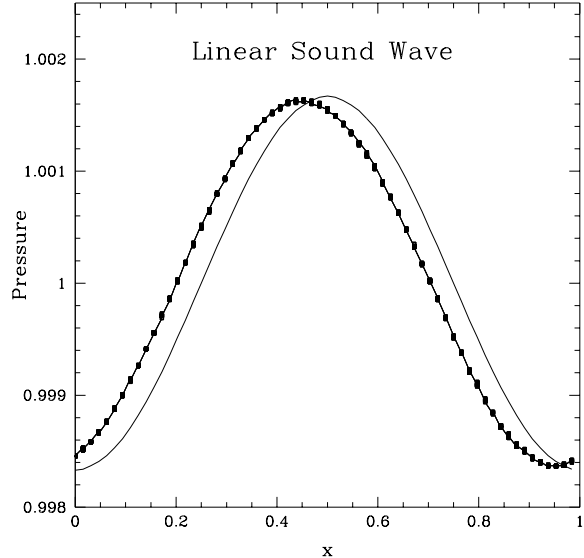} \hfill}
\caption{Second-order $64 \times 8 \times 1$ GPM result for the 
propagation of a right-moving linear sound wave with amplitude $\delta
v =0.1\%$ at time $t=10.0$.  The analytical result is the solid line
and the boxes indicate the GPM results with a connecting line to
assist in comparison.}
\label{fig:sndlin}
\end{figure}

Figure~\ref{fig:sndlin} shows the second-order GPM result for the
propagation of a linear acoustic wave with a sinusoidal profile.
Particles were placed on a regular grid in a periodic box of size $1.0 \times
0.125 \times 0.015625$ using $64 \times 8 \times 1$ particles.  The
initial conditions for a single eigenmode moving to the right were
imposed with a velocity perturbation of $\delta v =0.1\%$. The
smoothing length was fixed at $h=0.0488496$ and the CFL fraction was
$0.0125$ using the loose CFL condition.  Artificial viscosity was
employed with $\alpha=0.05$ and $ \beta=0.1$.  The sound speed is
$c_s=1.0$, so the wave should repeat itself each $1.0$ time units; the
plot time is $t=10.0$.  
The discrepancy with respect to the analytical expectation can be
divided into amplitude and phase error: the amplitude error appears to
be related to the timestep used and the value of artificial viscosity
employed, and the phase error (error in wave propagation speed) is
reduced with increasing spatial resolution.

\begin{figure}[h!]
\hbox to \hsize{ \hfill \epsfxsize8cm \epsffile{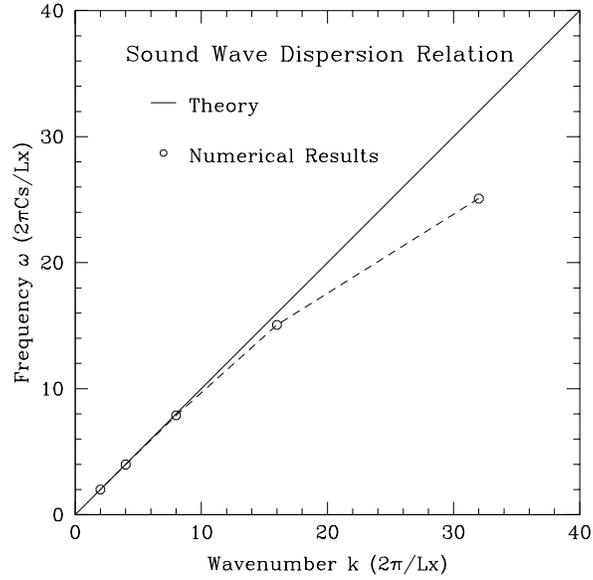} \hfill}
\caption{Dispersion relation for linear sound wave using the GPM algorithm.}
\label{fig:snddisp}
\end{figure}

Figure~\ref{fig:snddisp} presents the dispersion relation for linear
sound waves for varying wavenumber (or equivalently for varying number
of particles per wavelength).  All parameters are the same as the
previous test expect for: variable smoothing length is employed using
approximately 16 neighbors per particle ($\pm 33\%$); the CFL fraction
is $0.025$; and the resolution, or number of particles per wavelength,
is varied as indicated with corresponding changes in the box size.
The figure shows that for 32 or more particles per wavelength the GPM
results agree well with analytical predictions.

\begin{figure}[h!]
\hbox to \hsize{ \hfill \epsfxsize8cm \epsffile{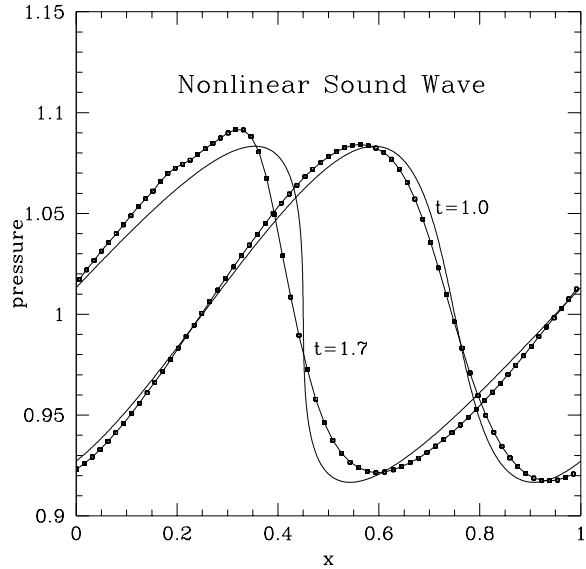} \hfill}
\caption{The propagation and steepening of a nonlinear sound wave of 
amplitude $\delta v=5.0\%$ using first order GPM with $64 \times 8
\times 1$ particles.  The analytical solution from an inviscid method
of characteristics is given by the solid line and the boxed line shows
the GPM results. The analytical solution forms a shock at $t=1.74$. }
\label{fig:sndnl}
\end{figure}

The steepening of a nonlinear sound wave into a shock is a sensitive
test of any hydrodynamical scheme.  Figure~\ref{fig:sndnl} shows the
GPM nonlinear sound wave compared to an inviscid method of
characteristics solution.
The parameters for this simulation are the same as the linear sound
wave above expect for: the amplitude of the perturbation is $\delta v=
5.0\%$, a variable smoothing length with approximately 16 neighbors
is used, and the calculation is done using first order GPM.
Analytically the formation of a shock occurs at $t=1.74$.  The GPM
results are plotted with analytical solutions for $t=1.0$ and $t=1.7$.
Because the artificial viscosity is quite low ($\alpha=0.05$,
$\beta=0.1$), this simulation is susceptible to large post-shock
oscillations; the beginnings of a post shock oscillation can be seen
in the GPM result at $t=1.7$.

\begin{figure}[h!]
\hbox to \hsize{ \hfill \epsfxsize8cm \epsffile{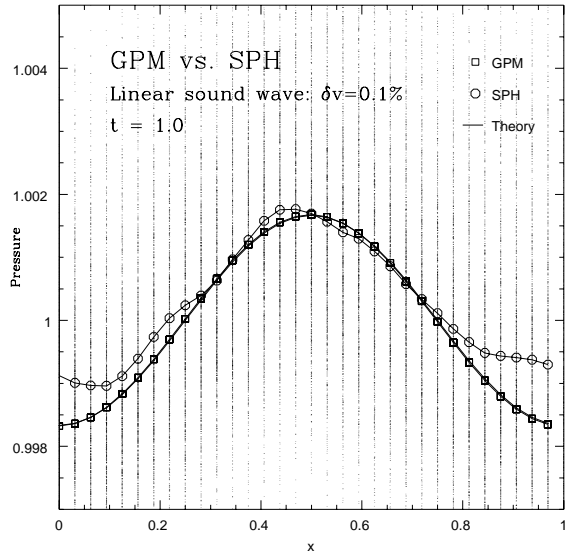} \hfill}
\caption{GPM vs. SPH results for a linear sound wave of amplitude $\delta
v=0.1\%$.  All SPH particles, which show a tremendous amount of
dispersion, are plotted as small dots and the average of all redundant
particles is shown as a circle.  The GPM particles are plotted as
boxes and show no dispersion.  The analytical result is the solid
sinusoidal line.}
\label{fig:sphlin}
\end{figure}

Figure~\ref{fig:sphlin} compares GPM results for a linear sound wave
with results from the SPH code Hydra \citep{cou95}.  The SPH code used
$32 \times 32 \times 32$ particles on a regular lattice in a periodic
box of size $1.0 \times 1.0 \times 1.0$.  Hydra is a publicly available
SPH code used primarily for cosmological simulations, so we choose to
turn off as many of the advanced options as possible: gravity,
expansion, and cooling were shut off.  The sound speed was set to
$c_s=1.0$ and a variable smoothing length is used with 32 neighbors
specified.  We used a simple but rigorous nearest neighbor search for
GPM which employs of order $N^2$ computations so, due to computational
limitations, the first-order GPM calculations used $32 \times 8 \times 8$
particles on a regular lattice in a periodic box of size $1.0 \times
0.25 \times 0.25$.  Artificial viscosity was turned off, the sound
speed was $c_s=1.0$, variable smoothing lengths were used with 32
neighbors, and the CFL fraction was $0.025$ according to the initial
particle separation.  Both codes were initialized with a single sound
wave propagating in the $x$ direction with amplitude $\delta
v=0.1\%$. The results are compared at $t=1.0$ and for both codes all
particles are plotted.
Note that after one sound crossing time of the box, the individual SPH
particles have suffered a tremendous amount of cross-field dispersion.
This is a result of the substantial unphysical noise inherent in the
SPH scheme due to its Monte Carlo nature; only an average of many
particles will provide an accurate result.  GPM demonstrates no
unphysical dispersion.

\begin{figure}[h!]
\hbox to \hsize{ \hfill \epsfxsize8cm \epsffile{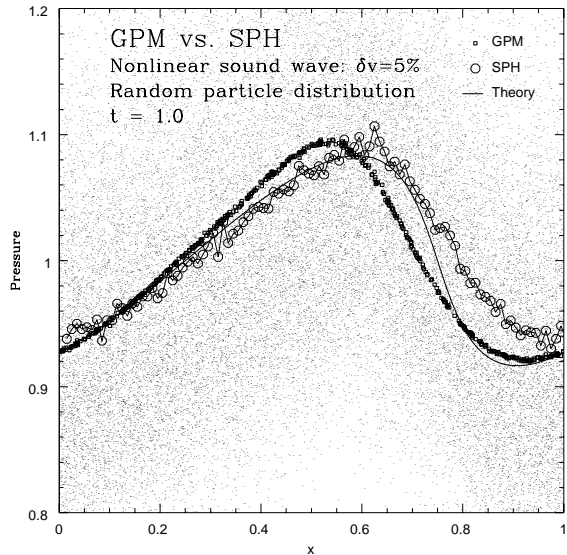} \hfill}
\caption{GPM vs. SPH results for a nonlinear sound wave of amplitude $\delta
v=5.0\%$.  Again, all particles for both methods are plotted;
demonstrating the large dispersion of SPH and small dispersion of GPM.
Bin averages of width $\Delta x=0.01$ for SPH results are given by
circles.}
\label{fig:sphnl}
\end{figure}

GPM results for a nonlinear sound wave of amplitude $\delta v=5.0\%$
are compared to SPH results in Figure~\ref{fig:sphnl}.  All parameters
are the same as for the previous comparison except: the number of
neighbors for GPM variable smoothing length was 45; and, for SPH,
$32^3$ particles are placed randomly in the box and, for GPM, $32
\times 8 \times 8$ particles are placed randomly.
Again, the large dispersion of SPH particles is clearly seen; only
averages of many particles, taken over all particles within bins of
width $\Delta x=0.01$, provide a solution that resembles the
analytical result. GPM shows no dispersion and hence requires no
averaging to yield an accurate solution; this encouraging result
suggests that the effective resolution of GPM is significantly higher
than that of SPH because there is no need to average over a large
number of particles to eliminate unphysical noise and obtain accurate
results.
 
\subsection{MHD Waves}

\begin{figure}[h!]
\hbox to \hsize{ \hfill \epsfxsize8cm \epsffile{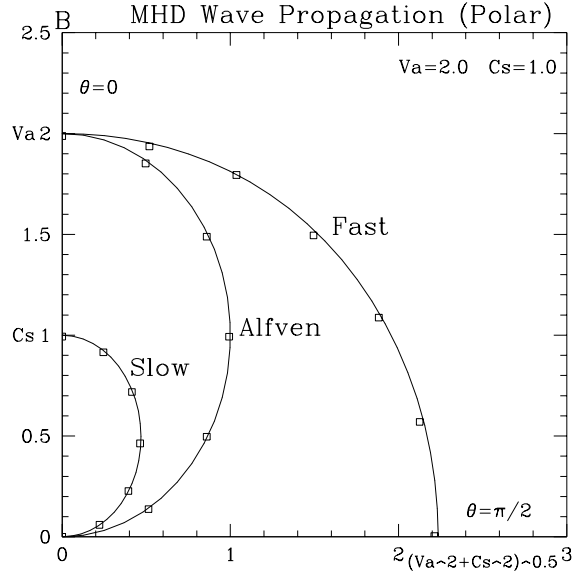} \hfill}
\caption{ Polar plot of the MHD wave speeds vs. the angle between the 
magnetic field $\bb$ and the wave propagation direction $\kk$.  The
analytical solutions for slow, \Alfven, and fast MHD waves are
indicated by the solid lines; second order GPM results for $32 \times
5\times 5$ simulations are given by the boxes. }
\label{fig:mhddisp}
\end{figure}

To test the ability of the GPM algorithm to accurately simulate MHD
phenomena, simulations of slow, \Alfven, and fast MHD waves were
performed over the full range of angles between the wave propagation
direction $\kk$ and the direction of the unperturbed magnetic field
$\bb$.  A complete discussion of their phase speeds and mode
eigenvectors is given in Appendix~\ref{alfven}.  The results of these
tests are easily summarized on a polar plot of MHD linear wave
propagation as shown in Figure~\ref{fig:mhddisp}.  In this plot, the
direction of the magnetic field is along the ordinate and the angle
between the magnetic field $\bb$ and the wave propagation direction
$\kk$ is the polar angle from the ordinate to the abscissa.  The
radius of the polar coordinate corresponds to the magnitude of the
wave velocity.  The analytical solutions are plotted as solid lines
and the boxes represent values obtained by the GPM code.  The second
order GPM simulations were run with $32 \times 5\times 5$ particles on
a uniform lattice in a periodic box of size $1.0 \times 0.15625 \times
0.15625$.  The smoothing length was fixed at $h=0.0508$ and artificial
viscosity parameters of $\alpha=0.05$ and $\beta=0.1$ were applied.
The sound speed was $c_s=1.0$, the \Alfven speed was $v_A=2.0$, and
the angles between $\bb$ and $\kk$ were $0^o$, $15^o$,$30^o$,
$45^o$,$60^o$, $75^o$, and $90^o$. The CFL fraction was $0.0125$ using
the loose CFL constraint.  A single eigenmode for each of the three
wave types was used with the amplitude of the velocity perturbation
$\delta \mathbf{v} \cdot \hat{\mathbf{k}}=0.1\%$; the direction of
propagation is along the $x$-axis for all simulations.
The GPM algorithm gives an excellent agreement with theory for all
three MHD waves over the entire range of propagation directions.

\begin{figure}[h!]
\hbox to \hsize{ \hfill \epsfxsize8cm \epsffile{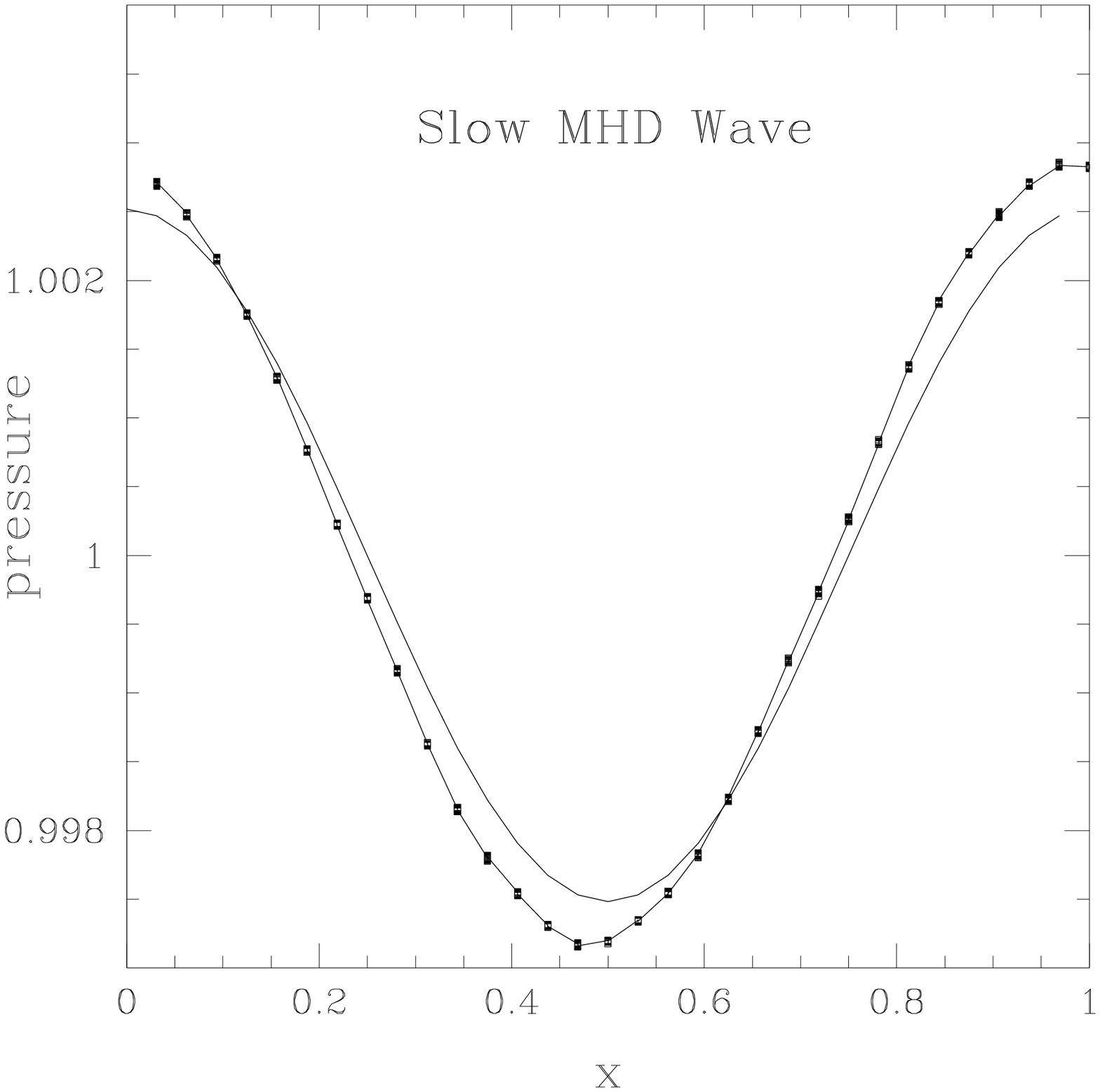} \hfill
\epsfxsize8cm \epsffile{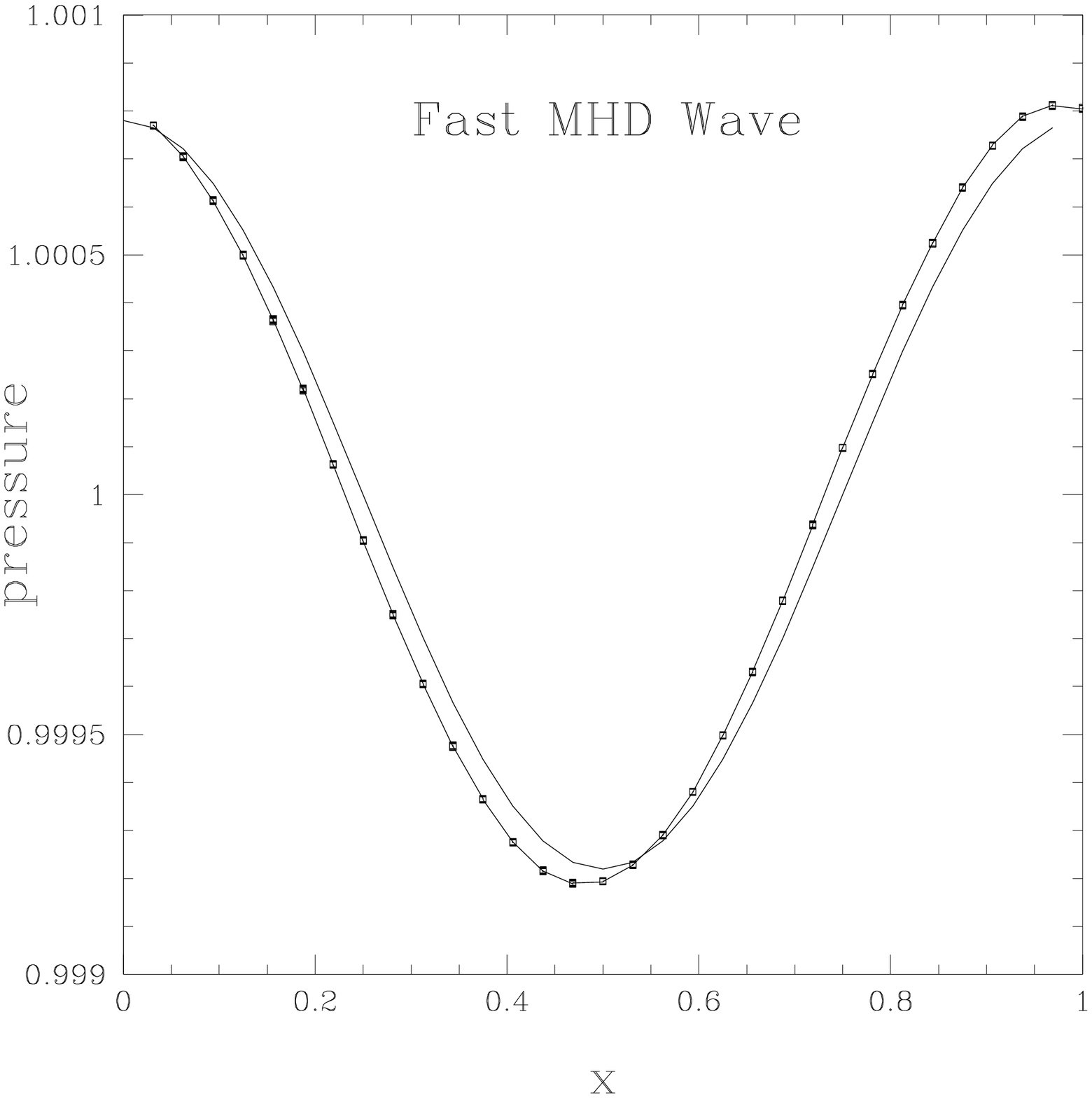} \hfill}
\caption{ Slow MHD wave (left) and fast MHD wave (right) propagating 
at $\theta=45^o$ from the magnetic field. Plots are shown at three
crossing times of the periodic box for that wave mode, $t=4.53$ for
the slow wave and $t=1.40$ for the fast wave.  Analytical results are
given by the solid line and second order GPM results with $32 \times 5
\times 5$ particles are given by the boxed line.}
\label{fig:slowfast}
\end{figure}

\begin{figure}[h!]
\hbox to \hsize{ \hfill \epsfxsize8cm \epsffile{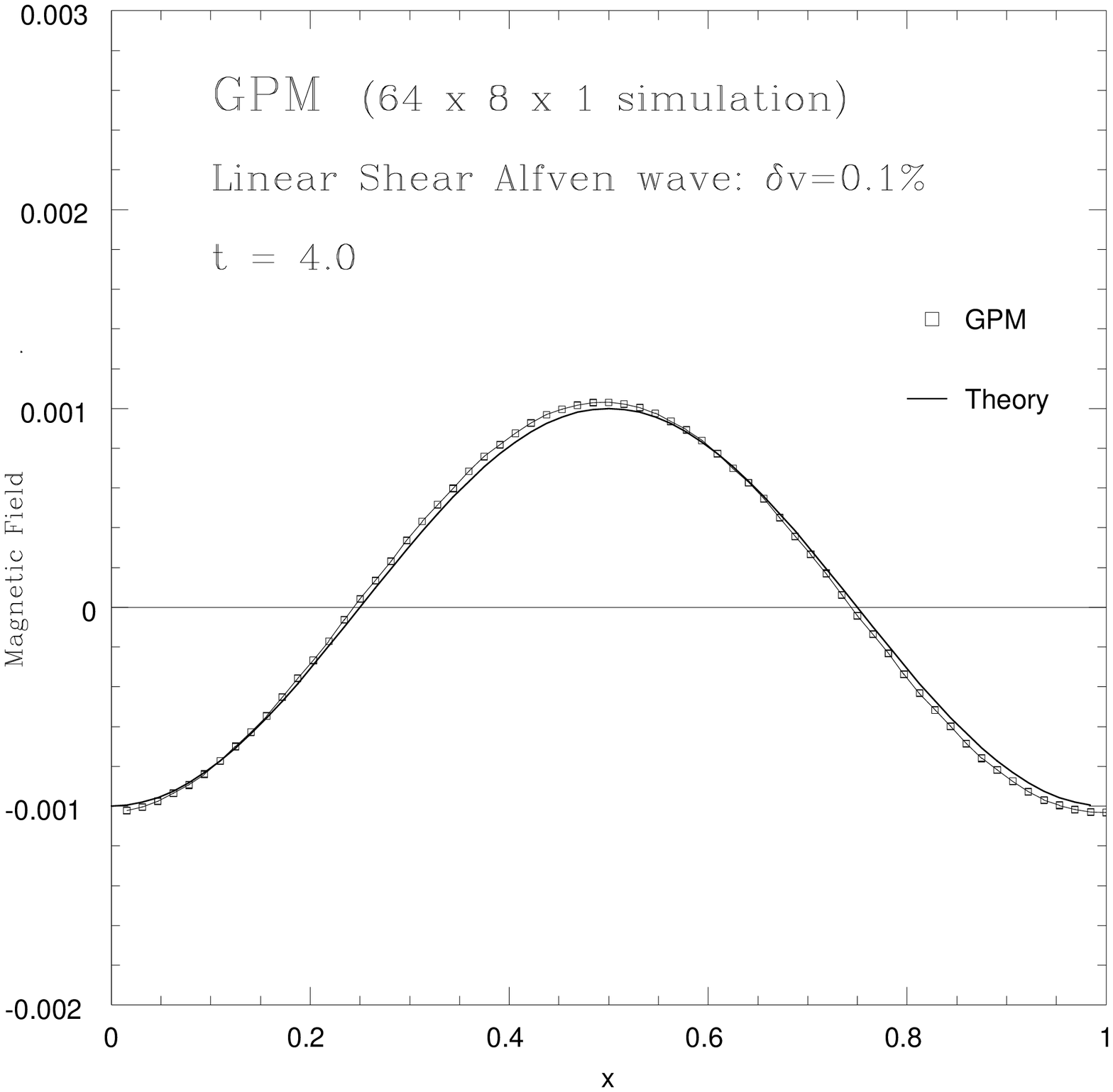} \hfill
\epsfxsize8cm \epsffile{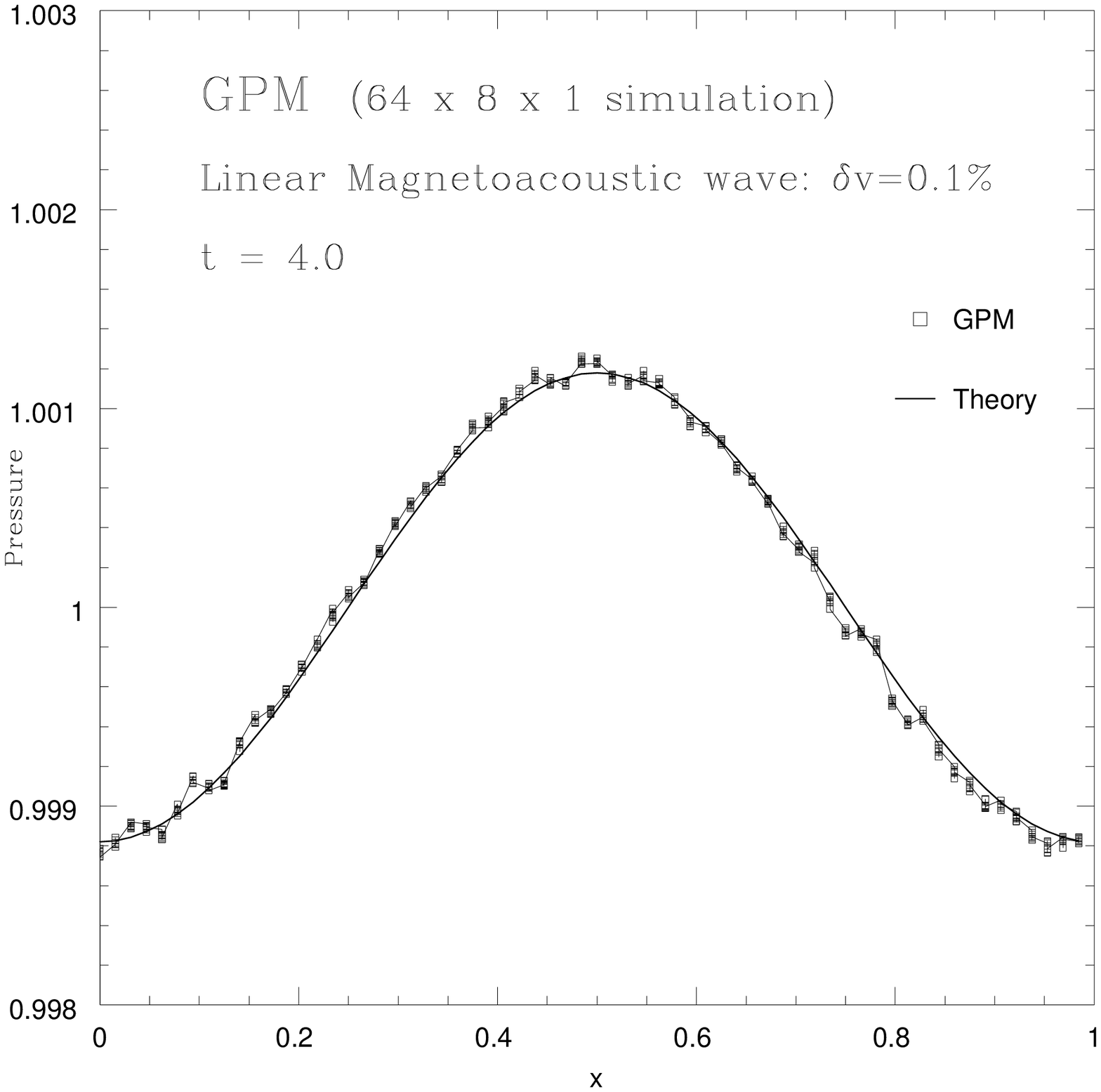} \hfill}
\caption{ Shear \Alfven wave propagating parallel to the magnetic field 
(left) and a magnetoacoustic wave propagating transverse to the
magnetic field (right). Plots are shown at three crossing times of the
periodic box for that wave mode, $t=1.5$ for the shear \Alfven wave
and $t=1.342$ for the magnetoacoustic wave. Analytical results are
given by the solid line and second order GPM results with $32 \times 5
\times 5$ particles are given by the boxed line.}
\label{fig:shearmagacoust}
\end{figure}

Examples of several of the simulations summarized in
Figure~\ref{fig:mhddisp} are shown in Figures~\ref{fig:slowfast} and
\ref{fig:shearmagacoust}. A slow MHD wave and a fast MHD wave both
propagating at $\theta=45^o$ from the magnetic field are shown in
Figure~\ref{fig:slowfast} for the third repetition time in the
periodic box (corresponding to $t=4.53$ for the slow wave and $t=1.40$
for the fast wave).
Figure~\ref{fig:shearmagacoust} shows a shear \Alfven
wave propagating at $\theta=0^o$ and a magnetoacoustic wave (fast wave
at $\theta=90^o$); both are plotted for the third repetition in the
periodic box, corresponding to $t=1.5$ and $t=1.342$, respectively.

\subsection{Advective MHD Problem: 2-D Magnetized Vortex}
\begin{figure}[h!]
\hbox to \hsize{ \hfill \epsfxsize8.2cm \epsffile{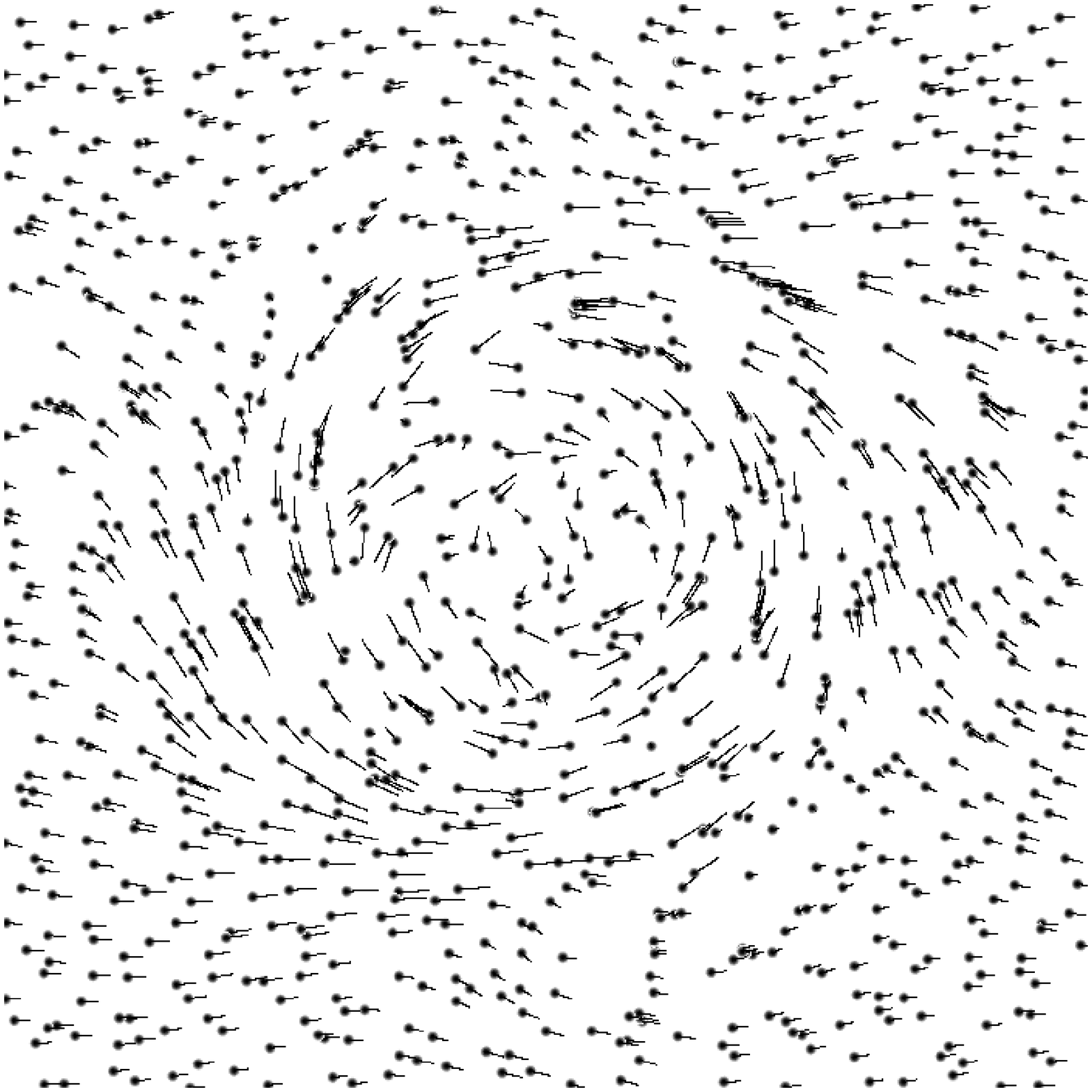} \hfill 
\epsfxsize8.2cm \epsffile{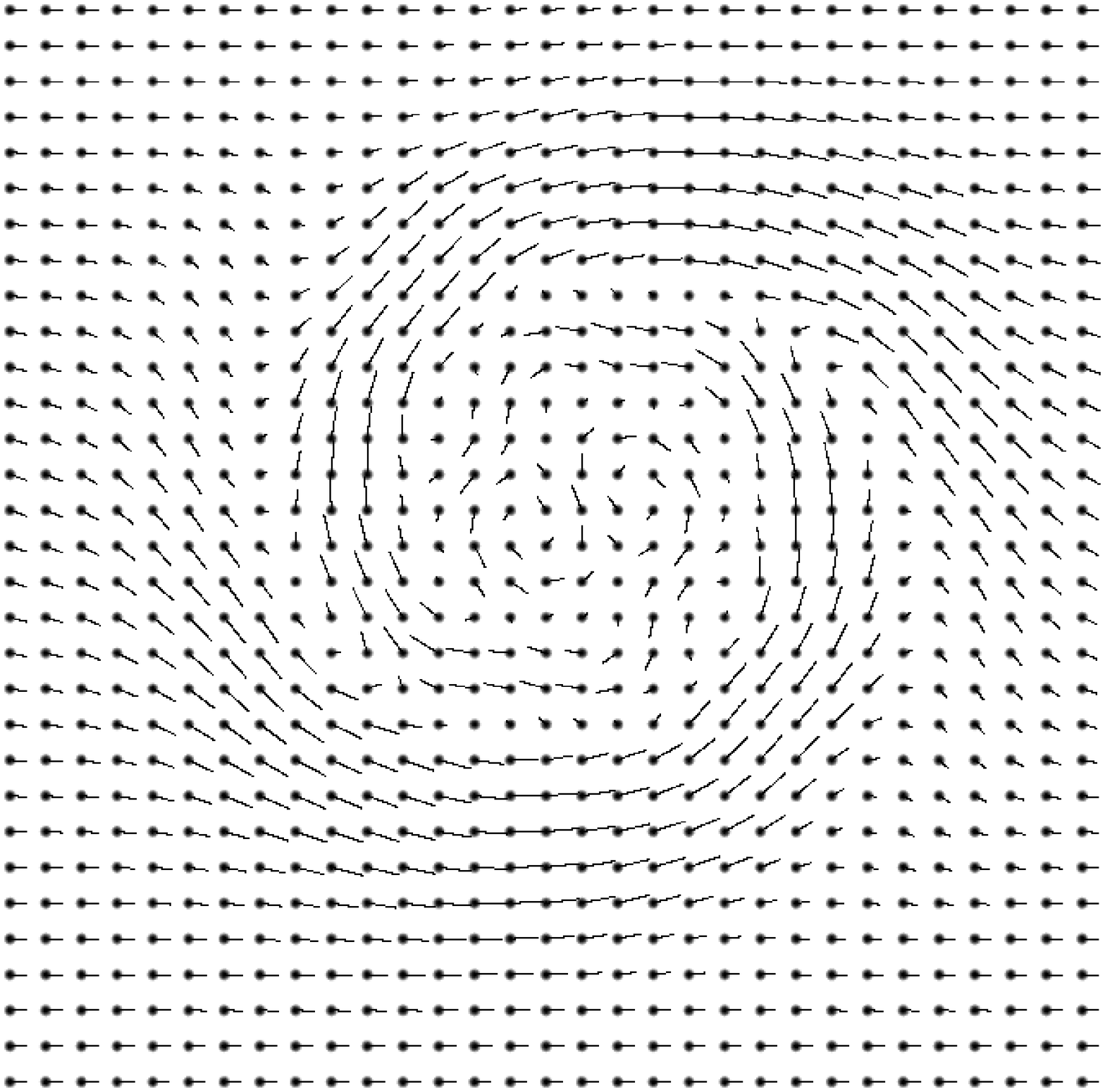} \hfill}
\caption{ GPM (left) and spectral MHD (right) simulation results 2-D 
magnetized vortex.  The spheres represent particle positions and the
arrows represent magnetic fields. A spectral code does not have
particles, and so the particles shown here serve only as markers for
the magnetic field arrows. }
\label{fig:bvortex}
\end{figure}

To test the GPM evolution of a dynamical magnetic field, we simulated
a 2-D vortex flow superimposed with an initially uniform magnetic
field.  The flow is initialized with an azimuthal flow profile of the form
\begin{equation}
v_\phi = v_0 \frac{r}{r_0} e^{(1-r^2/r_0^2)}
\end{equation}
with the values $v_0=0.1$ and $r_0=0.1667$ in a 2-D periodic box of
size $1.0 \times 1.0$.  The initial weak magnetic field is $\bb=0.001
\hat{\mathbf{x}}$.  Second order GPM is used with a fixed smoothing
length $h=0.123$ and artificial viscosity parameters $\alpha=0.05$ and
$\beta=0.1$.  The $32^2$ points are placed on a pseudorandom grid and
the CFL fraction is $0.0125$ assuming all particles are separated by a
distance $s=0.03125$.  The sound speed $c_s=1.0$.  The radius at the
peak of the azimuthal velocity will have undergone one full rotation
in a time $t= 10.47$.  Figure~\ref{fig:bvortex} shows the GPM results
at time $t=10.0$.  
For comparison, we also simulated the vortex with a spectral MHD code
\citep{mar01} and found good agreement with the GPM result. For the
GPM calculation, the magnetic field evolution is stable and magnetic
structures are resolved to 2 interparticle radii. In fact, the effective
viscosity of GPM is almost as good as that of the spectral simulation.

In a purely hydrodynamic version of this simulation using SPH, which
is not shown, the vortex stops turning after roughly one quarter
rotation due to intrinsic diffusivity of the SPH method. The GPM
vortex has not slowed appreciably.

\subsection{Shocks}

\begin{figure}[h!]
\hbox to \hsize{ \hfill \epsfxsize6cm \epsffile{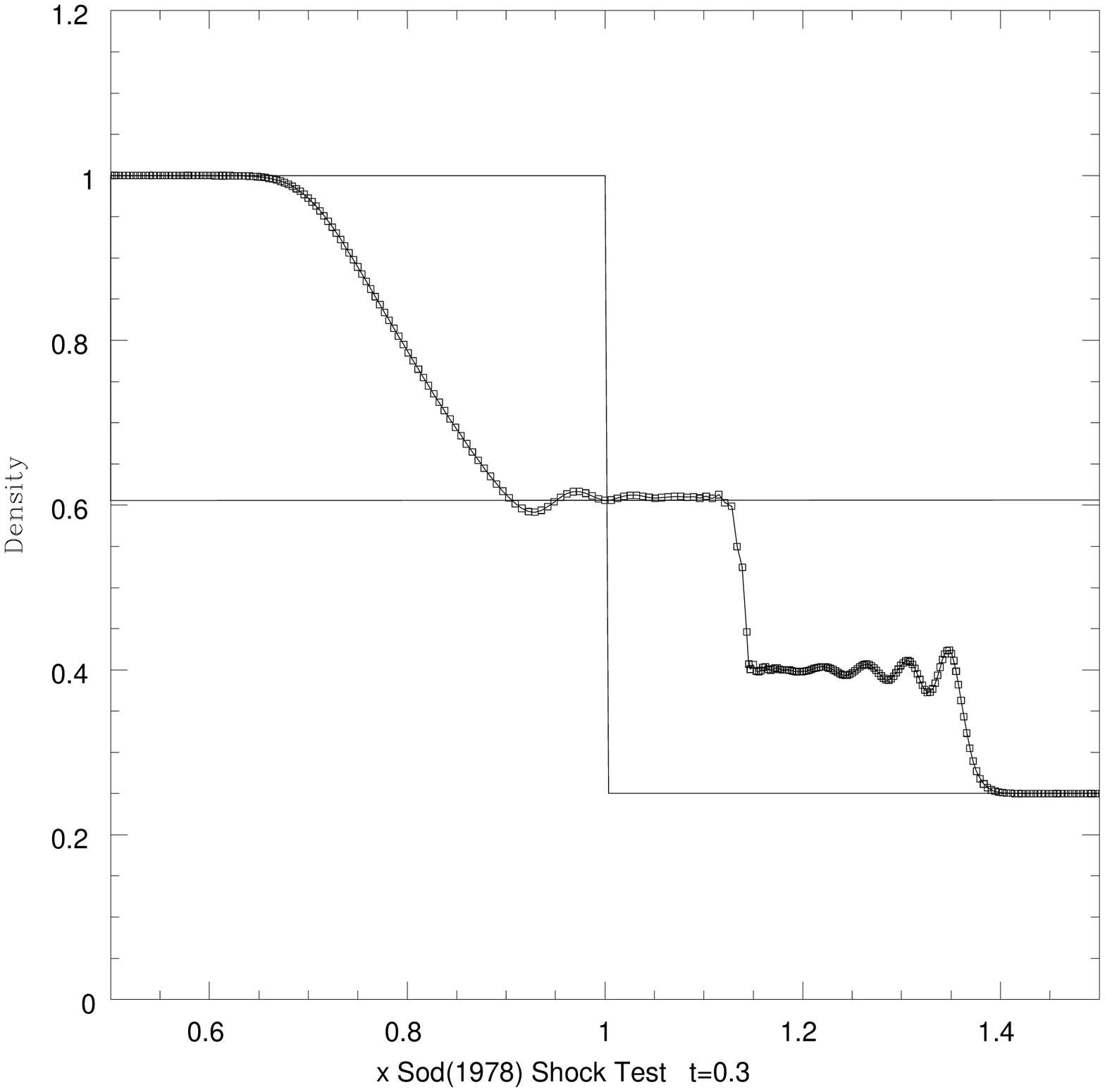} \hfill \epsfxsize6cm
\epsffile{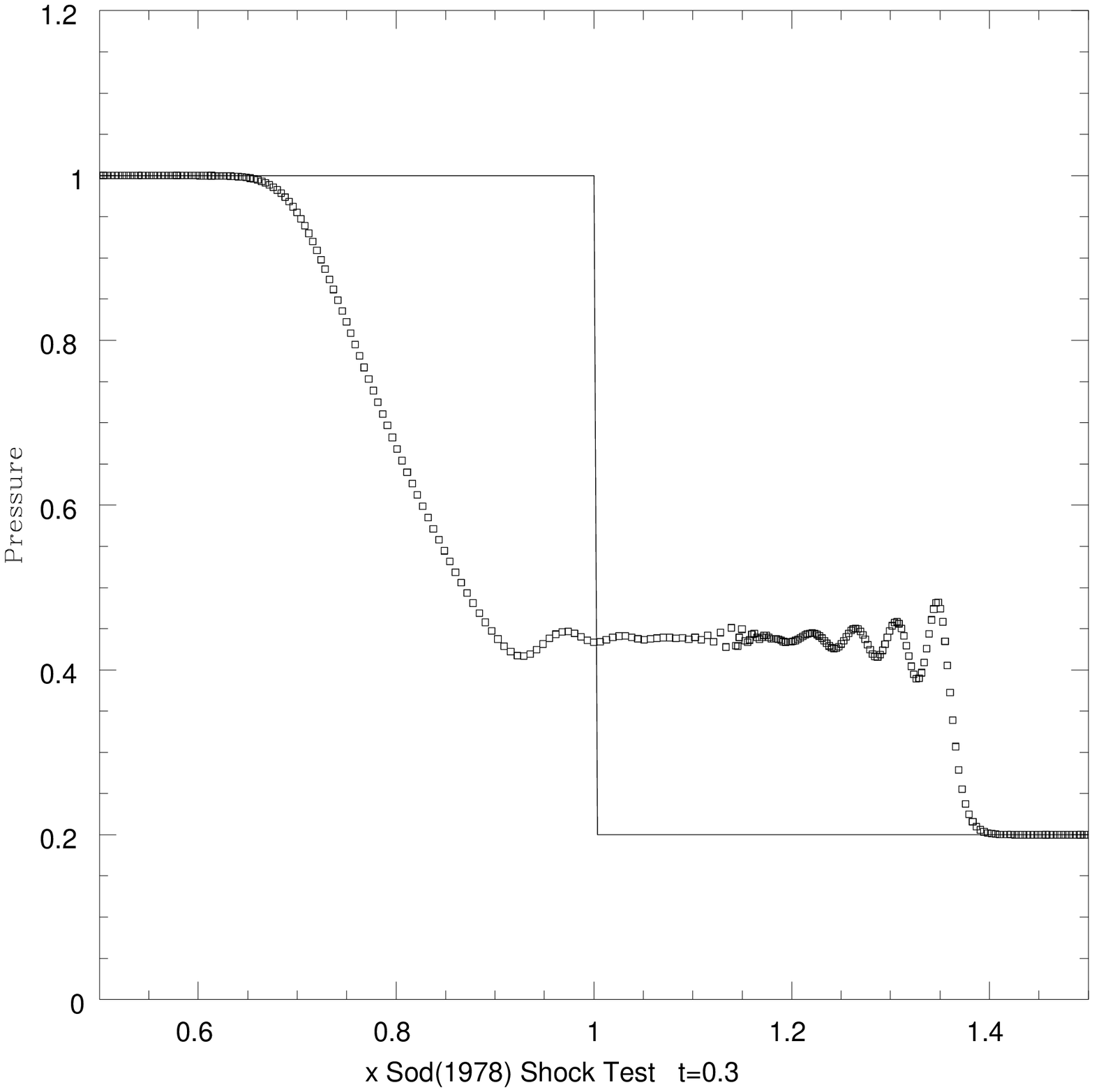} \hfill}
\hbox to \hsize{ \hfill \epsfxsize6cm \epsffile{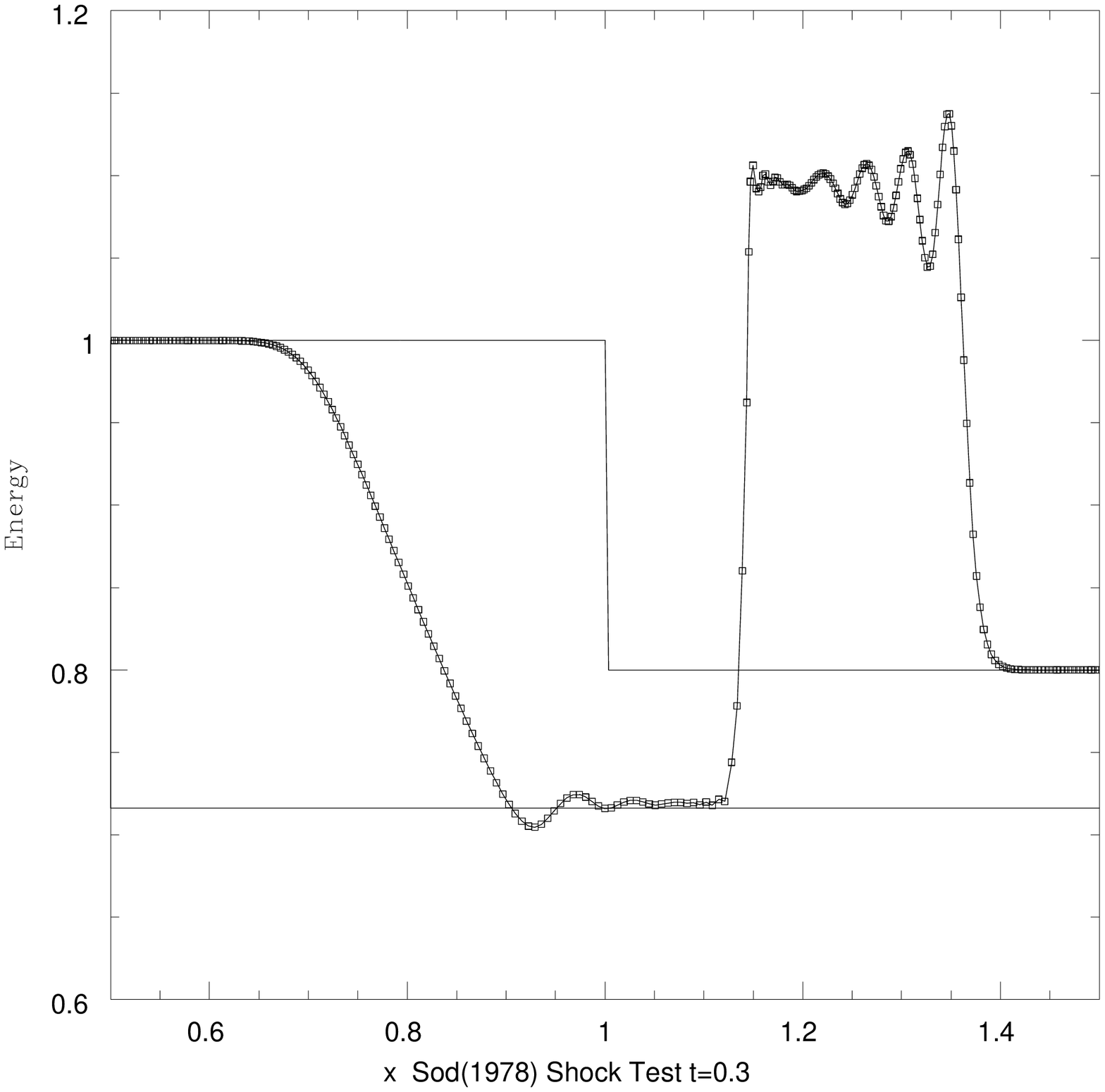} \hfill
\epsfxsize6cm \epsffile{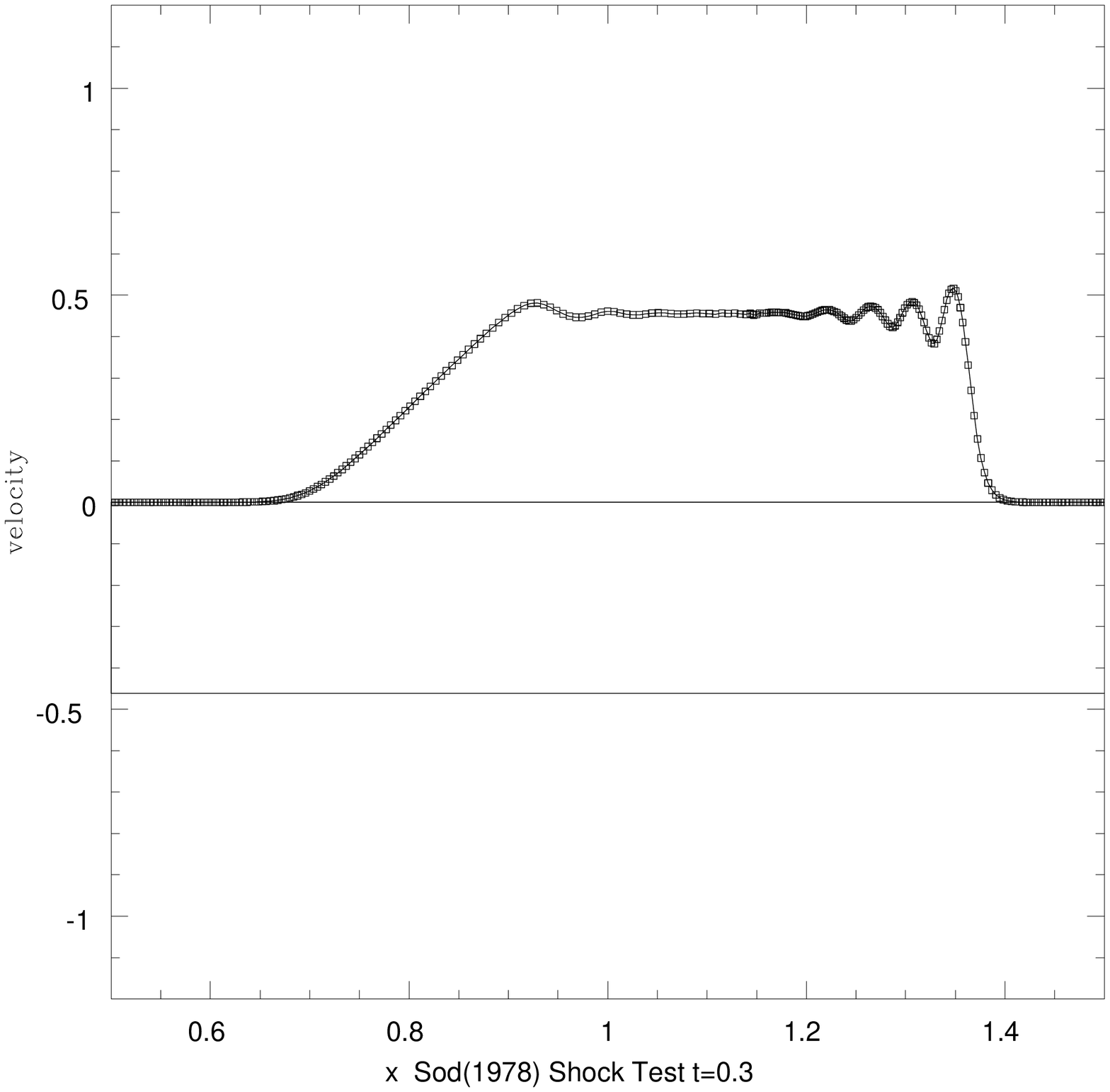} \hfill}
\caption{ The profile results of the standard \citet{sod78} 1-D shock 
test for density (upper left), pressure (upper right), energy (lower
left), and $x$-component of velocity (lower right) for the time
$t=0.3$.  Note the contact discontinuity visible in the density and
$x$ velocity profiles. }
\label{fig:sod}
\end{figure}

To test the shock-capturing capabilities of GPM, we used the standard
\citet{sod78} 1-D shock test.  Although quantities vary only in one
dimension, these tests were conducted in more than one dimension to
insure that there is no unphysical cross-field dispersion into the
redundant dimensions.  This stringent test begins with an initial
pressure and density discontinuity at an interface; compression and
rarefaction waves propagate into either side of the interface with a
contact discontinuity visible in the density and energy profiles only.
We choose the same initial conditions as the \citet{sod78} paper:
$p=1.0$ and $\rho=1.0$ to the left of the discontinuity, $p=0.1$ and
$\rho=0.125$ to the right, and zero velocity everywhere.  And, for
this problem only, we employ the adiabatic index $\gamma=1.4$ to
retain consistency with the original paper.  The resulting profiles
for density, pressure, energy, and $x$-component of the velocity are
shown in Figure~\ref{fig:sod} for time $t=0.3$.

\subsection{Collapse Problem}
To test the multiscale capability of the GPM code, we choose to do a
3-D collapse problem in a fixed gravitational potential and
demonstrate the attainment of a state of hydrostatic equilibrium.  

\section{Discussion}
\label{sec:disc}
The validation tests which have been presented here have all been
performed using a first-order Eulerian time stepping scheme.  To
achieve computational stability, we observed the timestep constraint
to be very tight: in most cases, the fraction of the
Courant-Friedrichs-Lewy limit was $0.0125$ to prevent any growth in
the amplitude of a linear wave.  A value of $0.025$ was also used and,
although the amplitude was seen to grow slightly (a few percent) over
a very long time ( around 100 sound crossing times of the periodic
box), it did not cause any problems.  A higher-order time stepping
scheme may help this significantly, but a tighter timestep constraint
may not be a problem if the GPM algorithm proves to have far superior
spatial resolution compared to SPH.

In the comparisons presented above with SPH results, GPM appears to be
able to resolve more detail with an equivalent number of particles
than SPH.  Because SPH is a Monte Carlo method, the error in the value
of a fluid quantity at any point scales as $N^{-1/2}$ for a random
distribution of particles and as $N^{-1}$ for a pseudo-random
distribution.  Therefore, to resolve the quantity to high precision, a
number of points must be averaged.  But GPM gives the value of the
fluid quantity at each point to the error of the chosen order of the
method, so there is no need for an averaging.  The spatial averaging
of any quantity necessarily involves a reduction in the effective
resolution of the method.  Thus,a GPM simulation may indeed
demonstrate a significant gain in resolution over a simulation in SPH
with an equivalent number of particles.  And, GPM is capable of stably
handling MHD, while SPH must resort to various tweaks of the method to
model MHD effects.

Another advantage of GPM over SPH, which cannot be extended beyond
a first-order scheme, is that GPM can, in theory, determine fluid
profiles to a specified polynomial degree.  In practice, the
limitations of computational power may prevent use of the method
beyond second order.  A first-order calculation requires the inversion
of a $4 \times 4$ matrix, a second-order a $10 \times 10$ matrix, a
third oder calculation a $20 \times 20$ matrix, and so on.  Hence, a
third-order calculation is likely to be too computationally expensive
to justify the higher-order determination.  But, even if one only
utilizes first-order GPM, it is useful to have a higher-order
extension of the method for convergence tests.

In the application of GPM presented here, we do not couple the mass
density of the fluid with the number density of ``particles'' used in
the calculation.  Hence, GPM is not a particle method in the
traditional sense; it behaves more like a deforming mesh of points at
which we know information about the fluid.  This prevents large
unphysical fluctuations in the density and allows the method to be
extended to higher order. It also simplifies the setting of initial
conditions in that the number density of points does not have to be
proportional to the local mass density.  As well, resolution can be
enhanced in a region of interest by simply placing more particles
there without disturbing the behavior of the fluid.  The disadvantage
of doing this is that it is possible, over long periods of time, for
mass density to slip away from the particles; conversely, it is
possible for the particles to gather in some areas, leaving voids in
other areas where the fluid properties are undersampled.  A method of
particle removal where an excess of particles have gathered and
particle replacement where voids leave the fluid undersampled can
rectify this potential problem.  In the simulations presented here, we
did not find any problems of this nature.

The other disadvantage of Lagrangian methods in general is the
difficulty of ensuring strict conservation of quantities such as mass,
linear and angular momentum, energy, and magnetic flux.  Further tests
of GPM will need to be carried out to determine the conservation
properties of GPM and the magnitude of any errors that arise.

Because GPM is very non-dissipative, discontinuities established in
initial conditions must be smoothed before beginning the simulation.
This is accomplished by using a first-order GPM calculation, either
with fixed or variable smoothing length, to determine the average
$\overline{q}$ of a fluid quantity $q$ and finding the smoothed value
$q_{sm}$ using $q_{sm}= q + f_{sm} (\overline{q} - q)$.  In practice,
one desires a smoothing length large enough so that a shock
discontinuity is spread out over at least $4$ particles.  

When performing a simulation with an external potential, as in the 3-D
collapse problem above, we found that it is critical for the form of
the applied force to be well sampled by particles.  If this is not
ensured, then the GPM algorithm cannot balance the applied external
forces with the internal fluid forces.  It is wise, also to smooth out
the gravitational force so as not to have any harsh discontinuities in
its form.  We found that a smoothed external gravitational force of
the form
\begin{equation}
F_g(r)= - \frac{GM}{r^2} \left( 1 - e^{-(r/r_0)^3} \right) \hat{\mathbf{r}}
\label{eq:smgrav}
\end{equation}
worked very well.

Finally, when doing isolated problems, we consider the question of the
ability of GPM to correctly determine forces at boundaries.  If there
are enough particles within the smoothing sphere, GPM seems to do a
reasonably good job of determining the forces.  In practice, it seems
best that the gradient goes to zero at the boundaries so that all
forces will go to zero. Thus, it is a good idea to place particles far
enough out that the forces are indeed nearly zero.  Since the mass
density is not related to particle number density, this can be done
with a very sparse distribution of particles.

\section{Conclusion}
\label{sec:conc}

The past two decades have seen extensive use of Lagrangian numerical
methods to model astrophysical phenomena.  SPH has proven to be an
invaluable tool for astrophysics to tackle problems with other
numerical methods simply could not handle.  But, as we move into the
next decade, an improvement in the precision of methods is needed to
refine existing theoretical models and to investigate problems in new
regimes.  In the absence of a viable extension of SPH to include
magnetic effects, astrophysicists have turned to grid-based methods
employing AMR to study MHD problems with large density contrasts.  In
this environment, we present Gradient Particle Magentohydrodynamics
(GPM) as an alternate Lagrangian method which accurately and stably
simulates MHD phenomena and which potentially can yield a significant
improvement in spatial resolution over SPH.  Various algorithms
employed in astrophysical computation are compared in
Appendix~\ref{app:algo}.

The simple recipe for the GPM algorithm is presented here and its
application to develop an MHD simulation code is described.  The GPM
scheme can be extended to higher orders and is observed to be very
non-diffusive.  We have inclusion validation tests to show the
behavior of GPM in modeling linear and nonlinear sound waves, the
full suite of MHD waves, an advective MHD problem, shocks, and
three-dimensional collapse.  These tests demonstrate clearly the
promise of this technique, although there is certainly room for
refinement of the method.

Although the two schemes are quite closely related in spirit, we
believe GPM to be superior to the SPH scheme for computing fluid
forces.  The similarity of the two numerical methods---that both
simply need a list of the nearest neighbors and the values of the
fluid quantities at those particles---makes the implementation of GPM
highly attractive.  Any existing high performance SPH code could
easily be modified to employ the GPM algorithm as the heart for the
determination of MHD forces instead of the SPH formulation.

Currently, the development of AMR for grid-based techniques is
extending the capability of grid-based codes into the regime
heretofore dominated by Lagrangian codes.  But, for the computational
power available today, AMR is still limited to a small number of
grid-refinement steps (in the neighborhood of 5 to 10 levels of
refinement).  As well, grid-based codes are susceptible to upwinding
and transport diffusion, and those incorporating AMR even more so.  A
Lagrangian code based on GPM is an ideal complement to the
computational effort of AMR codes.  The strengths of Lagrangian codes
are the weaknesses of grid codes, and vice-versa.  But, as
computational power increases, both numerical techniques should
converge to the same solution from different regimes: AMR from the
high-precision, more diffusive side with a smaller resolved range of
densities, and GPM from the low-precision, less diffusive side with a
greater resolved range of density.

\acknowledgements
We wish to thank Steve Cowley and James McWilliams for useful
discussions. We used the SPH code Hydra by \citet{cou95}.

\appendix

\section{Comparison of algorithms}
\label{app:algo}
There are 5 major classes of hydro/MHD algorithms, each having
strengths and weaknesses. A spectral code computes gradients in
Fourier space with the use of the Fast Fourier Transform (Borue \&
Orszag 1996, Maron \& Goldreich 2001). A grid code computes mass and
momentum fluxes through grid-cell boundaries. An example is the 3D
Compressible MHD code ``Zeus", a well validated program (Stone \&
Norman, 1992 I, 1992 II) enjoying widespread use in the astrophysical
community. An adaptive grid (Berger \& Oliger 1984, Berger \& Colella
1989) additionally allows cells to be dynamically subdivided in areas
of need. SPH (Monaghan 1992, Couchman, et. al., 1995) is smoothed
particle hydrodynamics and GPM is gradient particle
magnetohydrodynamics.

\begin{center} {\it Table \tablemethod: The principal fluid dynamics
algorithms.  Items 1-4 refer to physics capabilities of the
algorithms. Items 6-9 refer to matters of computational efficiency.}
\end{center}

\begin{center}
\begin{tabular}{lccccc}
\hline
Characteristic         & Spectral & Grid & Adaptive grid & SPH   & GPM\\
\hline
1) Subsonic            &   Yes    &  Yes & Yes  & Noisy &      Yes   \\
2) Supersonic (Shocks) &   No     &  Yes & Yes  & Yes   &      Yes   \\
3) Magnetic fields     &   Yes    &  Yes & Yes  & No    &      Yes   \\
4) Resolution          &   Best   &  Good& Good & Poor  &      Good  \\
5) Code complexity     &   Easy   &  Easy& Hard & Easy  &      Easy  \\
6) Lagrangian timesteps&   No     &  No  & No   & Yes   &      Yes   \\
7) Spatial refinement  &   No     &  No  & Yes  & Yes   &      Yes   \\
8) Variable timesteps  &   No     &  No  & Yes  & Yes   &      Yes   \\
9) Parallelizable      &   Yes    &  Yes & Yes  & Yes   &      Yes   \\
\hline
\end{tabular}
\end{center}

\begin{center} {\it Table \tablemethodz: Computational efficiency} \end{center}
\begin{center}
\begin{tabular}{ll}
\hline
Subsonic:            &The velocities are less than the sound speed.\\
Supersonic:          &The velocities exceed the sound speed, producing\\
                     &strong density contrasts and shocks.\\
Magnetic fields:     &The ability to include magnetic fields.\\
Resolution:          &The effective resolution per computational element.\\
                     &This property is distinct from spatial refinement.\\
Code complexity:     &The algorithmic complexity of the code.\\
Lagrangian timesteps:&Timesteps are determined by local velocity\\
                     &fluctuations instead of the local average velocity.\\
                     &Therefore, larger timesteps can be used, increasing\\
                     &computational efficiency.\\
Spatial refinement:  &The ability to focus resolution on areas of interest.\\
Variable timesteps:  &The ability to use smaller timesteps in areas of need\\
                     &while the rest of the system can simultaneously evolve\\
                     &with a larger timestep.\\
Parallelizable:      &The ability to run on several processors \\
                     &simultaneously, increasing execution speed.\\
\hline
\end{tabular}
\end{center}

\section{SPH}

SPH simulations serve as a reference for several of our validation tests.
Standard SPH is discussed in the ARAA review by Monaghan and has
seen extensive refinement since. It is based on the smoothed spatial
average, here for a general quantity $q$, of the form:
\begin{equation}
q = h^{-3} \sum_b \frac{m_b q_b}{\rho_b} W(r).
\end{equation}
where the subscript $b$ denotes neighboring particles. $W(r)$ is a symmetric
smoothing kernel with a characteristic radius $h$, and $\nabla W(r)$
is the gradient kernel.
If $q$ is the density, then
\begin{equation}\rho = h^{-3} \sum_b m_b W(r).\end{equation}

The pressure operator is symmetrized to conserve momentum.
The subscript $a$ refers to the test particle.
\begin{equation}
-\frac{1}{\rho} \nabla P
= - \nabla \frac{P}{\rho} - \frac{P}{\rho^2} \nabla \rho
= - h^{-3} \sum_b m_b
    \left( \frac{P_b}{\rho_b^2} + \frac{P_a}{\rho_a^2} + \Pi \right)
    \nabla W \end{equation}
An artificial viscosity has been added through an extra pressure term $\Pi$
that acts only between converging particles.
\begin{equation}
\Pi = \frac{-\alpha c_{ab} \mu_{ab} + \beta \mu_{ab}^2}{\rho_{ab}}
\hspace{.cm} ; \hspace{.3cm} v \cdot r < 0 \hspace{1.5cm}
\Pi = 0 \hspace{.cm} ; \hspace{.3cm} v \cdot r > 0 \end{equation}

\begin{equation}
\rho_{ab} = \frac{\rho_a + \rho_b}{2} \hspace{1.5cm}
c_{ab} = \frac{1}{2} \left(
         \frac{\gamma_a P_a}{\rho_a} + \frac{\gamma_b P_b}{\rho_b} \right)
\hspace{1.5cm}
\mu_{ab} = \frac{h (v_b - v_a) \cdot r}{r^2 + .01 h^2} \end{equation}

The density and energy equations have similar forms.
\begin{equation}
d_t \rho = - \rho \nabla \cdot v = h^{-3} \sum_b m_b (v_a-v_b)
\cdot \nabla_a W \end{equation}
\begin{equation}
d_t e = - \frac{P}{\rho} \nabla \cdot v
= \frac{1}{2} h^{-3} \sum_b m_b
  \left( \frac{P_b}{\rho_b^2} + \frac{P_a}{\rho_a^2} \right) (v_a - v_b)
  \cdot \nabla_a W
\end{equation}

\section{MHD waves} \label{alfven}

We summarize here the MHD eigenvectors for arbitrary
Alfv\'en and acoustic speeds $v_A$ and $v_S$ (Shu 1992), assuming only
that the fluctuations are small with respect to the Alfv\'en and
acoustic speeds.  The uniform magnetic component has a value of $b_0$
oriented along $\hat{\z}$. Define mean and fluctuating quantities as
$$
b = b_0  \hat{\z} + b_0 (b_{\hat{a}} \hat{\aaa} + b_{\hat{s}} \hat{\s}) \hh
e^{i(\kk\cdot\x - \omega t)} \hspace{8mm}
v = (v_{\hat{k}} \hat{\kk} + v_{\hat{a}} \hat{\aaa} + v_{\hat{s}} \hat{\s}) \hh
e^{i(\kk\cdot\x - \omega t)}
$$
$$
\rho = \rho_0 + \rho_0\rho_1 \hh e^{i(\kk\cdot\x-\omega t)} \hspace{5mm}
$$
where the unit right-hand polarization vectors $\hat{\kk}$,
$\hat{\s}$, and $\hat{\aaa}$
are defined by
\begin{equation}
\hat{\bf a}\equiv {\hat{\kk}\times\hat{\z}\over
[1-(\hat{\kk}\cdot\hat{\z})^2]^{1/2}}, \hspace{12mm} \hat{\bf s}\equiv
{\hat{\z}-(\hat{\kk}\cdot\hat{\z})\hat{\kk}\over
[1-(\hat{\kk}\cdot\hat{\z})^2]^{1/2}}.
\label{eq:polarization}
\end{equation}
and $\cos \theta = \hat{\kk} \cdot \hat{\z}$.

The
Alfv\'en wave dispersion relation is $\omega^2/k^2 = v_A^2 (\hat{\z}
\cdot \hat{\kk})$, and the eigenvectors are
$$\rho_1 : v_{\hat{a}} : b_{\hat{a}} \hh = \hh 0 : \pm v_A : -1. $$
The dispersion relations for the
fast (+) and slow (-) modes are $$ \frac{w^2}{k^2} =
\frac{1}{2} \left[ (v_A^2 + v_S^2) \pm [ (v_A^2 + v_S^2)^2
      - 4 v_A^2 v_S^2 \cos^2 \theta ]^{1/2} \right] $$
The eigenvectors are
$ \rho_1 : v_{\hat{k}} : v_{\hat{s}} : b_{\hat{s}} $
$$
= \hh \cos \psi \sin \psi : \frac{w}{k} \cos \psi \sin \psi :
\frac{w}{k} \left[ \sin^2 \psi + \frac{v_S^2 - w^2/k^2}{v_A^2} \right] :
- \frac{v_S^2 - w^2/k^2}{v_A^2} \cos \psi. $$

Acoustic waves are a special case where there is no magnetic field.
The progate at the acoustic speed and have an eigenvector of $\rho_1 :
v_{\hat{k}} = 1 : v_S$.

GPM accurately reproduces the wave speeds and eigenvectors for all
three polarizations of linear waves for all propagation directions.
\section{Application of Courant-Friedrichs-Lewy Timestep Constraint}
There are two basic implementation of the CFL time control, loose and
strict.  Consider the Courant-Friedrichs-Lewy stability criterion that
\[ \frac{|v| \Delta t}{\Delta x} \le 1. \]
We can determine the timestep to use by 
\[ \Delta t = f_{CFL} \frac{\Delta x}{|v|} \]
where $f_{CFL}$ is the Courant fraction applied to the problem.

The loose implementation finds the mean nearest neighbor particle
separation $\overline{s}$ and the standard deviation of that value
$\sigma_s$. It employs the value $\Delta x= \overline{s}-\sigma_s$ and
$|v| = \mathrm{max}(v_f,v_{i(max)})$ where $v_f$ is the fast wave
speed and $v_{i(max)}$ is the fastest particle speed in the system.

The strict implementation calculates the value $\frac{\Delta x}{|v|}$
for each particle and its nearest neighbor where $\Delta x$ is the
distance to the nearest neighbor and $|v| =
\mathrm{max}(v_f,v_{rel})$.  Again, $v_f$ is the fast wave speed and
$v_{rel})$ is the relative velocity between the two particles.  It
selects the minimum value of $\frac{\Delta x}{|v|}$ in the entire
system and calculates the timestep as above.

Note that, contrary to what the names of the two implementations
suggest, for a system with a roughly uniform particle distribution,
the loose implementation can actually be a more stringent control on
the timestep than the strict implementation if the fastest particle in
the system is used as $|v|$.  The strict implementation is absolutely
necessary when a system has particle separations that vary
dramatically.



\begin{thebibliography}{}
\bibitem[Berger and Oliger(1984)]{ber84}
  M.J. Berger and J. Oliger, ``Adaptive mesh refinement for hyperbolic
  partial differential equations", J. of Comput. Phys., 53, 1984
\bibitem[Berger and Colella(1989)]{ber89}
  M.J. Berger and P. Colella, ``Local adaptive mesh refinement for shock
  hydrodynamics", J. Comp. Phys., 82, 64-84, 1989
\bibitem[Borue and Orszag(1996)]{bor96}
  Borue, V. \& Orszag, S. A., 1996, J. Fluid Mech., 306, 293.
\bibitem[Couchman et. al.(1995)]{cou95}
  Couchman, H. M. P., Thomas, P. A., Pearce, F. R., ``Hydra: an Adaptive-Mesh
  Implementation of P3M-SPH", 1995, ApJ, 452, 797
\bibitem[Lucy(1977)]{luc77} Lucy, ? 1977
\bibitem[Maron and Goldreich(2001)]{mar01}
  Maron, J. L. \& Goldreich, P. M., ``Simulations of Incompressible MHD
  Turbulence, 2001, ApJ.
\bibitem[Monaghan and Gingold(1977)]{mon77} Monaghan, J.J. and Gingold, R.A. 1977, ?

\bibitem[Monaghan and Gingold(1983)]{mon83} Monaghan, J.J. and Gingold, R.A. 1983, J. Comp. Phys., 52, 374
\bibitem[Monaghan(1985)]{mon85} Monaghan, J.J. 1985, Comp. Phys. Rep., 3, 71
\bibitem[Monaghan(1992)]{mon92}
  Monaghan, J. J., ``Smoothed Particle Hydrodynamics", 1992, ARAA, 30, 453
\bibitem[Roache(1975)]{roa75} Roache, P. J. 1975, Computational Fluid Dynamics.
 	Albequerque: Hermosa.
\bibitem[Shu(1992)]{shu92}
  Shu, F., ``The Physics of Astrophysics II, Gas Dynamics", 1992 
\bibitem[Sod(1978)]{sod78}
  Sod, G. A. 1978, Journal of Computational Physics, 27, 1-31
\bibitem[Stone and Norman(1992a)]{sto92a}
  Stone, J. \& Norman, M., ``Zeus-2D, A Radiation Magnetohydrodynamics Code
  for Astrophysical Flows in Two Space Dimensions: I. The Hydrodynamic
  Algorithms and Tests", 1992a, ApJS, 80, 753.
\bibitem[Stone and Norman(1992b)]{sto92b}
  Stone, J. \& Norman, M., ``Zeus-2D, A Radiation Magnetohydrodynamics Code
  for Astrophysical Flows in Two Space Dimensions: II. The Magnetohydrodynamic
  Algorithms and Tests", 1992b, ApJS, 80, 791.


\end{thebibliography}
\end{document}